# Dynamic control of ferroionic states in ferroelectric nanoparticles

Anna N. Morozovska[4,*], Sergei V. Kalinin[2,†], Mykola E. Yelisieiev[3], Jonghee Yang[4], Mahshid Ahmadi[4], Eugene. A. Eliseev[5], and Dean R. Evans[6,‡]

[1] Institute of Physics, National Academy of Sciences of Ukraine, 46, pr. Nauky, 03028 Kyiv, Ukraine

[2] The Center for Nanophase Materials Sciences, Oak Ridge National Laboratory, Oak Ridge, TN 37831, USA

[3] Taras Shevchenko National University of Kyiv, Volodymyrska street 64, Kyiv, 01601, Ukraine

[4] Institute for Advanced Materials and Manufacturing, Department of Materials Science and Engineering, University of Tennessee, Knoxville, TN, 37996, USA

[5] Institute for Problems of Materials Science, National Academy of Sciences of Ukraine, Krjijanovskogo 3, 03142 Kyiv, Ukraine

[6] Air Force Research Laboratory, Materials and Manufacturing Directorate, Wright-Patterson Air Force Base, Ohio, 45433, USA

**Abstract**

The polar states of uniaxial ferroelectric nanoparticles interacting with a surface system of electronic and ionic charges with a broad distribution of mobilities is explored, which corresponds to the experimental case of nanoparticles in solution or ambient conditions. The nonlinear interactions between the ferroelectric dipoles and surface charges with slow relaxation dynamics in an external field lead to the emergence of a broad range of paraelectric-like, antiferroelectric-like ionic, and ferroelectric-like ferroionic states. The crossover between these states can be controlled not only by the static characteristics of the surface charges, but also by their relaxation dynamics in the applied field. Obtained results are not only promising for advanced applications of ferroelectric nanoparticles in nanoelectronics and optoelectronics, they also offer strategies for experimental verification.

---

[*] corresponding author, e-mail: anna.n.morozovska@gmail.com
[†] corresponding author, e-mail: sergei2@ornl.gov, sergei2@utk.edu
[‡] corresponding author, e-mail: dean.evans@afresearchlab.com

## I. INTRODUCTION

Dimensionally confined ferroics have emerged as one of the objects of interest for the fundamental research and device applications [1]. In particular, ferroelectric thin films and nanoparticles are a playground for exotic physics driven by the interplay of polarization rotation, depolarization fields, surface charges [2, 3], flexoelectricity [4, 5], and size effects [6, 7]. For electrode surfaces and ferroelectric nanoparticles in a dielectric matrix, extensive theoretical work has been developed [8, 9, 10]. However, this is not the case for the prototypical case of ferroelectric nanoparticles in an ambient environment or in solutions. Here, despite a general understanding that the surface phenomena at open ferroelectric surfaces are dominated by electrochemical processes and ionic screening [11, 12, 13], their effects on physical properties and phase transitions in nanoferroics are understood only weakly [14]. For instance, ferroelectric thin films with open surfaces compensated via ionic adsorption [15, 16, 17], reveal very unusual mixed electrochemical-ferroelectric states [18, 19].

Previously, a theoretical formalism for the analysis of the ferroelectric behavior in proximity to electrochemically coupled interface was developed by Stephenson and Highland (**SH**) [20]. Using the Landau-Ginzburg-Devonshire (**LGD**) and Stephenson-Highland approaches, Morozovska and Kalinin groups derived analytical solutions describing unusual phase states in uniaxial [21, 22, 23, 24] and multiaxial [25] ferroelectric thin films, as well as antiferroelectric thin films with electrochemical polarization switching [26, 27]. The analysis [21-27] leads to the elucidation of ferroionic (**FEI**) and antiferroionic (**AFEI**) states, which are the result of a nonlinear electrostatic interaction between the surface charges (ions, vacancies, holes and/or electrons) and ferroelectric dipoles. These states have principal differences from the ferroelectric state of thin films with linear screening, which originate from the strong nonlinear dependence of the screening charge density on the acting electric potential (i.e., electrochemical "overpotential" - the electric potential inside the charged layer that is different from the external voltage due to the self-screening and depolarization field) [21]. This nonlinearity gives rise to additional multi-stable polarization states in ultrathin ferroelectric films. Comparatively, linear screening conditions can lead only to paraelectric (**PE**) states.

The FEI and AFEI states reveal themselves as having specific (e.g., nonmonotonic) temperature dependences of the out-of-plane spontaneous polarization and unusually complex shapes of the polarization hysteresis loops [21-27]. For instance, there can be four bi-stable states of the spontaneous polarization in an ultra-thin uniaxial ferroelectric film covered with ionic-electronic surface charges, as compared to two possible states in thicker films covered with conducting electrodes [21]. If the polarization hysteresis loop of the thin film covered with ions looks similar to a classical ferroelectric loop, we can classify the state as ferroelectric (**FE**)-like FEI, and when it looks similar to a double antiferroelectric loop, we can classify the state as AFEI. There are a plenty of mixed antiferroionic-ferroionic (**AFEI-FEI**) states characterized by e.g., pinched and/or constricted asymmetric loops. It should be noted that the appearance of AFEI and FEI states depends not only on the static characteristics surface charges (such as their concentration, formation energies, and the details of surface density of states), but also on their relaxation dynamics in an applied field. However, this dynamic aspect of ionic screening is almost unexplored [21-27], while ample experimental evidence of its existence is available [28, 29, 30].

Similarly, the influence of the nonlinear ionic-electronic surface screening on the polar properties of **ferroic nanoparticles** has not been considered, despite that this case seems very interesting for the fundamental research of dimensionally confined objects and promising for various applications of the nanoparticles in optoelectronics [31], ferroelectric random access memories [32], and other information technology components [33].

Traditionally, the influence of the surface screening on the polar properties of a single free-standing ferroelectric nanoparticle remains difficult to explore due to the lack of corresponding probing techniques. However, free-standing ferroelectric nanoparticles have become accessible to the experimental investigation with the advent of Piezoresponse Force Microscopy (**PFM**) [34] and other advanced probing techniques. For example, the existence and reordering behaviors of a surface dipole in a metal halide perovskite system has been observed by utilization of advanced probing techniques [35, 36]. The study provides a local, nanoscopic insight for understanding the ferroelectric properties of this class of materials. Nanoparticles of metal halide perovskites have recently come to the forefront of research due to their unique properties for variety of optoelectronic applications [37, 38]. A recent study revealed the existence of ferroelectricity in $CsPbBr_3$ nanostructure [39]. It is noted that such ferroelectric phases in the nanostructures can be stabilized and/or controlled by the class of ligand molecules with a wide-range of sizes, shapes, chirality, ionic characteristics, and polarity [40, 41]. Each ligand molecule can result in the surface system being identical to a model of nonlinear ionic-electronic surface screening; therefore, metal halide perovskite nanoparticles could be ideal systems to explore the ferroionic properties in the confined nanostructures.

At the same time, experimental realizations of ferroelectric nanoparticles placed in highly polarized nonlinear media, such as liquid crystals, are abundant (see e.g., Refs. [42, 43, 44]), and the sizes of 5 – 50 nm are typical experimental values [45, 46, 47, 48]. There are several studies of ferroelectric nanoparticles fabricated in heptane and oleic acid [49, 50, 51], producing core-shell nanoparticles, where nonlinear screening can be realized in the ultra-thin shells.

This theoretical study aims to fill the gap in knowledge and considers polar states and dielectric properties of uniaxial ferroelectric nanoparticles covered by a layer of mixed ionic-electronic nonlinear surface charge with slow relaxation dynamics in an external field. Using LGD-SH approach, we show that a dynamic transition between the PE-like, AFEI, FEI, and FE-like FEI states takes place in ultra-small ferroelectric nanocubes and nanopillars. The paper is structured as following. **Section II** contains the formulation of the problem and description of methods. Basic equations and assumptions are introduced in **Section III**. Results discussion and analysis are presented in **Section IV**, which consist of three subsections considering the electron-hole and electron-cation surface screening and size effects. A brief summary is given in **Section V**. Calculations details are listed in **Suppl. Mat**. [52].

## II. PROBLEM DESCRIPTION

The polarization, internal electric field, and elastic stresses and strains in ferroelectric nanoparticles of different shape, material, size, and structure are modelled using finite element modeling (**FEM**). In this, we use the constitutive equations for relevant order parameter fields based on the LGD phenomenological approach along with electrostatic equations and elasticity theory. For geometry, we consider a parallelepiped-shape nanoparticle of height $h$, which consists of a uniaxial ferroelectric core of thickness $h_f$ sandwiched between the dielectric layers of thickness $d_1$ and $d_2$, as shown in **Figs. 1(a)-(b)**; the dielectric permittivity of the layers is $\varepsilon_l$ and $\varepsilon_u$, respectively. The layers can have the same or different physical nature. Specifically, for PFM geometry, the upper layer can be gaseous, soft matter, or liquid gap between the ferroelectric core and the top (or tip) electrode; while the lower dielectric layer is an ultra-thin "passive" or "dead" layer [53, 54, 55] with high PE-type dielectric permittivity, which is formed at the interface between the ferroelectric core and conductive substrate. For this case, one can assume that $d_1 > d_2$ and $\varepsilon_l \gg \varepsilon_u$. The relative background permittivity [54] of a ferroelectric core is $\varepsilon_b$.

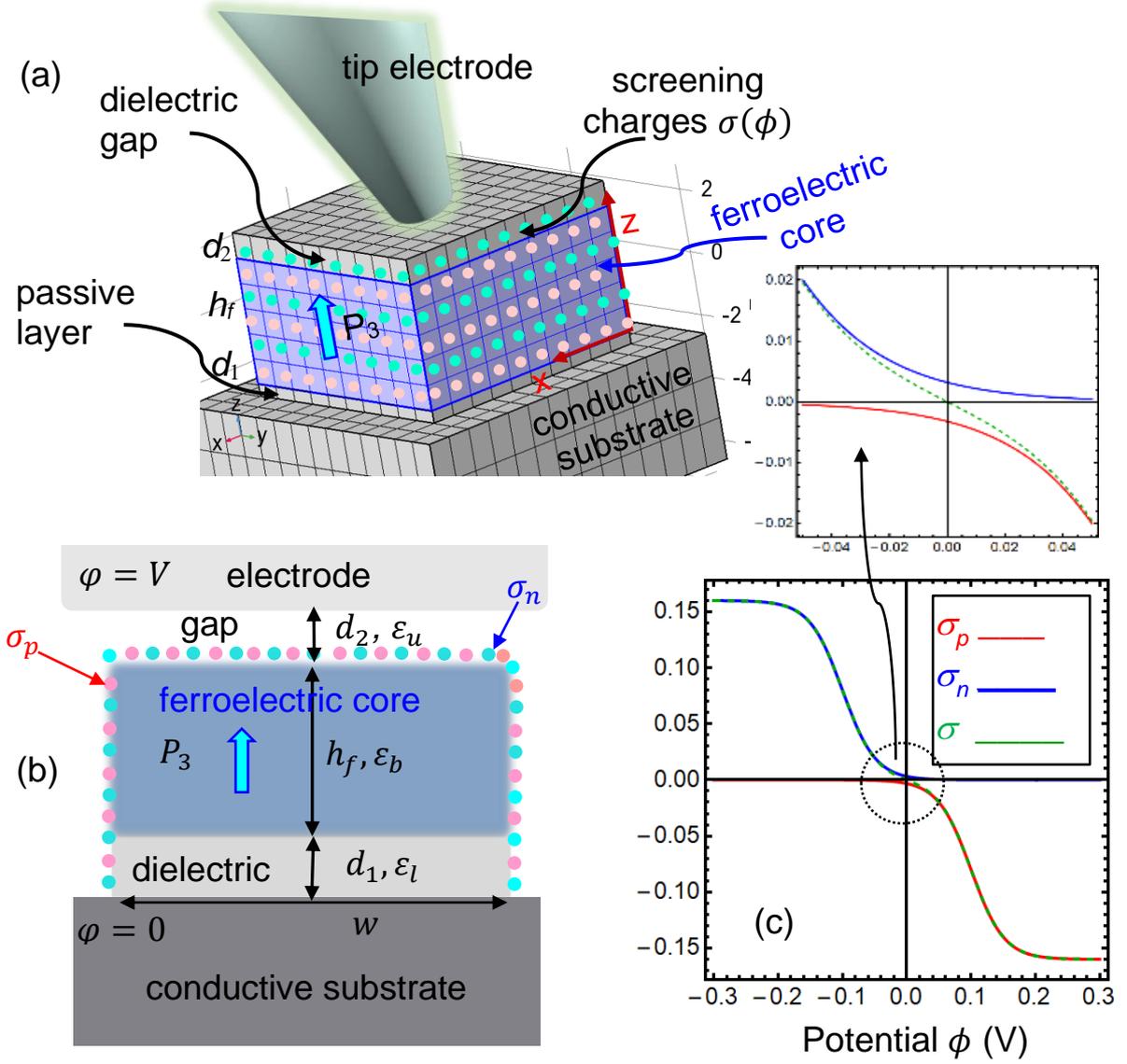

**FIGURE 1**. A parallelepiped-shape nanoparticle of height $h$, which consists of a uniaxial ferroelectric core of thickness $h_f$ sandwiched between two dielectric layers of thickness $d_1$ and $d_2$. The direction of the uniaxial polarization $P_3$ is shown by the blue arrow. The side view **(a)** and the vertical cross-section **(b)** are shown. The side and top surfaces of the ferroelectric core are covered with a layer of mobile surface charges with density $\sigma$, which is the sum of positive (pink circles) and negative (blue circles) charges, $\sigma_n$ and $\sigma_p$, respectively. **(c)** The dependence of the total surface charge $\sigma$ (green dashed curve), $\sigma_n$ (blue curve), and $\sigma_p$ (red curves) on the overpotential $\phi$, which are calculated from Eqs.(2) for $C_n = C_p = 10^{18}$ m$^{-2}$, $Z_p = -Z_n = +1$, $g_p = g_n = 1$, $\Delta G_p^0 = \Delta G_n^0 = 0.1$ eV, and $T = 298$ K. The inset shows the central part of the plot (c).

The z axis is parallel to the polar axis, and the uniaxial polarization $P_3 \uparrow\uparrow z$. The aspect ratio of the nanoparticle width $w$ to height $h$ is varied, such that the width ranges from being twice the height ($w \geq 2h$) to significantly greater than the height ($w \gg 2h$). This geometry variation corresponds to the transition from a nanocube to nanopillar shape. The nanoparticle bottom surface, $z = 0$, is in contact with an electron-conducting substrate; and the mismatch strain $u_m$ can exist at the nanoparticle-substrate boundary. The outer surface of the upper dielectric layer, $z = h$, is an electrode.

The top and side surfaces of the ferroelectric core are covered with a layer of mobile surface charges (holes and/or electrons, ions, protons, hydroxyl groups, etc.). The surface charge density $\sigma$ depends on the electric potential $\phi$ in a complex nonlinear way, where the total ionic-electronic charge is the sum of positive ($\sigma_p$) and negative ($\sigma_n$) charges, $\sigma(\phi) = \sigma_p(\phi) + \sigma_n(\phi)$, whose mobilities can vary significantly for different kinds of charges. This mobility difference can lead to the separable dynamics (retardation or outrunning) of $\sigma_n$ and $\sigma_p$. One cannot exclude the presence of a surface charge at bottom surface of the ferroelectric core, but its nature and properties can be strongly different from the ionic-electronic charge at the electrically-open top surface.

### III. BASIC EQUATIONS AND APPROXIMATIONS
#### A. Surface charge dynamics and LGD free energy

To describe the dynamics of the positive and negative surface charges, we use a linear relaxation model [21, 56]:

$$\tau_p \frac{\partial \sigma_p}{\partial t} + \sigma_p = \sigma_{p0}[\phi], \quad \tau_n \frac{\partial \sigma_n}{\partial t} + \sigma_n = \sigma_{n0}[\phi], \tag{1}$$

where the relation between the relaxation times $\tau_p$ and $\tau_n$ can be very different due to the dissimilar mobilities of the ionic and electronic charges $\sigma_p$ and $\sigma_n$. Typically, the relaxation of electrons and/or holes are of the same order, and, at the same time, the relaxation of electrons and holes are much faster than the relaxation of ions. Thus, below we consider that the equality $\tau_n \approx \tau_p$ is valid for the case when the surface charges are electrons and holes, while the strong inequalities $\tau_n \ll \tau_p$ or $\tau_p \ll \tau_n$ are valid for the case when the surface charges are electrons and cations, or holes and anions, respectively.

The dependence of the equilibrium charge densities, $\sigma_{p0}[\phi]$ and $\sigma_{n0}[\phi]$, on the potential $\phi$ is known, e.g., as proposed by Stephenson and Highland [13, 20]. In this work we will use the following extension of the SH model:

$$\sigma_{p0}[\phi] = eZ_p C_p \left(1 + g_p \exp\left(\frac{\Delta G_p^0 + eZ_p \phi}{k_B T}\right)\right)^{-1}, \tag{2a}$$

$$\sigma_{n0}[\phi] = eZ_n C_n \left(1 + g_n \exp\left(\frac{\Delta G_n^0 + eZ_n \phi}{k_B T}\right)\right)^{-1}, \tag{2b}$$

where $e$ is an elementary charge, $Z_{p,n}$ are the ionization degrees of the surface charges, and $C_{p,n}$ are the 2D surface charge concentrations (measured in m$^{-2}$). Positive parameters $\Delta G_p^0$ and $\Delta G_n^0$ are the free energies of the surface defects formation under normal conditions and zero potential ($\phi =$ 0). Exact values of $\Delta G_{p,n}^0$ are poorly known for many important cases, but are regarded as varying over the range (0.03 – 0.3) eV [13, 20]. Positive prefactors $g_p$ and $g_n$ can originate from different mechanisms of the charge formation [57, 58].

The electroneutrality condition equivalent to the total charge absence, $\sigma_{p0} + \sigma_{n0} = 0$, should be valid for the pairwise formation of negative and positive surface charges at $\phi = 0$, and the condition imposes limitations on the charge density parameters in Eqs.(2), namely:

$$\frac{Z_p C_p}{Z_n C_n} = -\frac{1 + g_p \exp\left(\frac{\Delta G_p^0}{k_B T}\right)}{1 + g_n \exp\left(\frac{\Delta G_n^0}{k_B T}\right)}. \tag{3}$$

Being interested in low frequency dynamics, below we impose condition (3). To fulfil this condition, we can consider the pairwise formation of negative and positive surface charges, when the charges have opposite signs, $Z_p = -Z_n = Z$, and their concentrations are equal $C_p = C_n = C$. Then, the following relation between the prefactors and formation energies, $\Delta G_p^0 - \Delta G_n^0 = k_B T \ln\left(\frac{g_n}{g_p}\right)$, should be valid. If the prefactors are equal, $g_p = g_n = g$, then the formation energies are equal too, $\Delta G_p^0 = \Delta G_n^0 = \Delta G$. A violation of condition (3) can take place far from equilibrium, e.g., under electrochemical reactions [13, 20]. The dependences of $\sigma$, $\sigma_n$, and $\sigma_p$ on the electric potential $\phi$ calculated from Eqs.(2) for typical characteristics of the surface charges are shown in **Fig. 1(c).**

Using condition (3) for small potential values, $\left|\frac{e Z_p \phi}{k_B T}\right| \ll 1$, we obtain that

$$\sigma_0[\phi] = -\varepsilon_0 \frac{\phi}{\lambda}, \tag{4a}$$

where we introduce the inverse effective screening length:

$$\frac{1}{\lambda} = \frac{(eZ)^2 C}{\varepsilon_0 k_B T} g \exp\left(\frac{\Delta G}{k_B T}\right)\left(1 + g \exp\left(\frac{\Delta G}{k_B T}\right)\right)^{-2}. \tag{4b}$$

The LGD free energy functional $G$ of a uniaxial ferroelectric includes a Landau energy – an expansion on 2-4-6 powers of the polarization component $P_3$, $G_{Landau}$; a polarization gradient energy, $G_{grad}$; an electrostatic energy, $G_{elect}$; an elastic, electrostriction, and flexoelectric contributions, $G_{elastic}$; and a surface energy, $G_S$. It has the form [59]:

$$G = G_{Landau} + G_{grad} + G_{elect} + G_{elastic} + G_S, \tag{5a}$$

$$G_{Landau} = \int_{V_f} d^3r \left(\frac{\alpha}{2} P_3^2 + \frac{\beta}{4} P_3^4 + \frac{\gamma}{6} P_3^6\right), \tag{5b}$$

$$G_{grad} = \int_{V_c} d^3r \frac{g_{33kl}}{2} \frac{\partial P_3}{\partial x_k} \frac{\partial P_3}{\partial x_l}, \tag{5c}$$

$$G_{elect} = -\int_{V_f} d^3r \left(P_i E_i + \frac{\varepsilon_0 \varepsilon_b}{2} E_i E_i\right), \tag{5d}$$

$$G_{es} = -\int_{V_f} d^3r \left( \frac{s_{ijkl}}{2} \sigma_{ij}\sigma_{kl} + Q_{ij33}\sigma_{ij}P_3^2 + \frac{F_{ij3l}}{2}\left(\sigma_{ij}\frac{\partial P_3}{\partial x_l} - P_3\frac{\partial \sigma_{ij}}{\partial x_l}\right)\right), \tag{5e}$$

$$G_S = \frac{1}{2}\int_S d^2r\, a_{33}^{(S)} P_3^2. \tag{5f}$$

Here $V_f = h_f S$ is the ferroelectric core volume. The coefficient α linearly depends on temperature $T$, $\alpha(T) = \alpha_T[T - T_C]$, where $\alpha_T$ is the inverse Curie-Weiss constant and $T_C$ is the ferroelectric Curie temperature renormalized by electrostriction and surface tension [60]. All other coefficients in Eqs.(5) are regarded as temperature-independent. The coefficient β is positive if the ferroelectric material undergoes a second order transition to the paraelectric phase and negative otherwise; and the coefficient $\gamma \geq 0$. The gradient coefficients tensor $g_{ijkl}$ are positively defined. In Eq.(5e), $\sigma_{ij}$ is the stress tensor, $s_{ijkl}$ is the elastic compliances tensor, $Q_{ijkl}$ is the electrostriction tensor, and $F_{ijkl}$ is the flexoelectric tensor.

### B. A single-domain state approximation

Since the stabilization of a single-domain polarization in ultrathin ferroelectric films takes place due to chemical switching (see e.g. Refs.[11, 12, 13]), we may assume that the distribution of the out-plane component $P_3(x, y, z)$ does not deviate significantly from the value $\bar{P}$ averaged over the nanoparticle volume, namely $P_3 \cong \bar{P}$.

The ranges of parameters for which the domain formation does not take place can be established by FEM. We performed FEM in COMSOL@MultiPhysics for $Sn_2P_2S_6$ (**SPS**) nanoparticles covered with an electronic-ionic charge with surface density given by Eq.(4a). Material parameters of SPS are listed in **Table SI** in the Supplement. The initial distribution of polarization was chosen randomly small.

It appeared that the ultra-small **SPS** nanocubes and nanopillars (with height between 2 and 20 nm) are either in a paraelectric (**PE**) or a single-domain ferroelectric (**SDFE**) state, which is dependent on the temperature and screening length $\lambda$. The polydomain ferroelectric state (**PDFE**) can be stable in bigger nanoparticles for $\lambda > \lambda_{cr}$. A typical phase diagram of the SPS nanopillars, which contains the regions of PE, SDFE, and PDFE are shown in **Fig. 2(a).** A typical polarization distribution inside a 4-nm nanocube is shown in **Figs. 2(b)-(d)** for several values of the effective screening length $\lambda$. The transition from the PE phase to the PDFE phase and then to the SDFE phase occurs with a decrease in $\lambda$ from 200 pm to 20 pm.

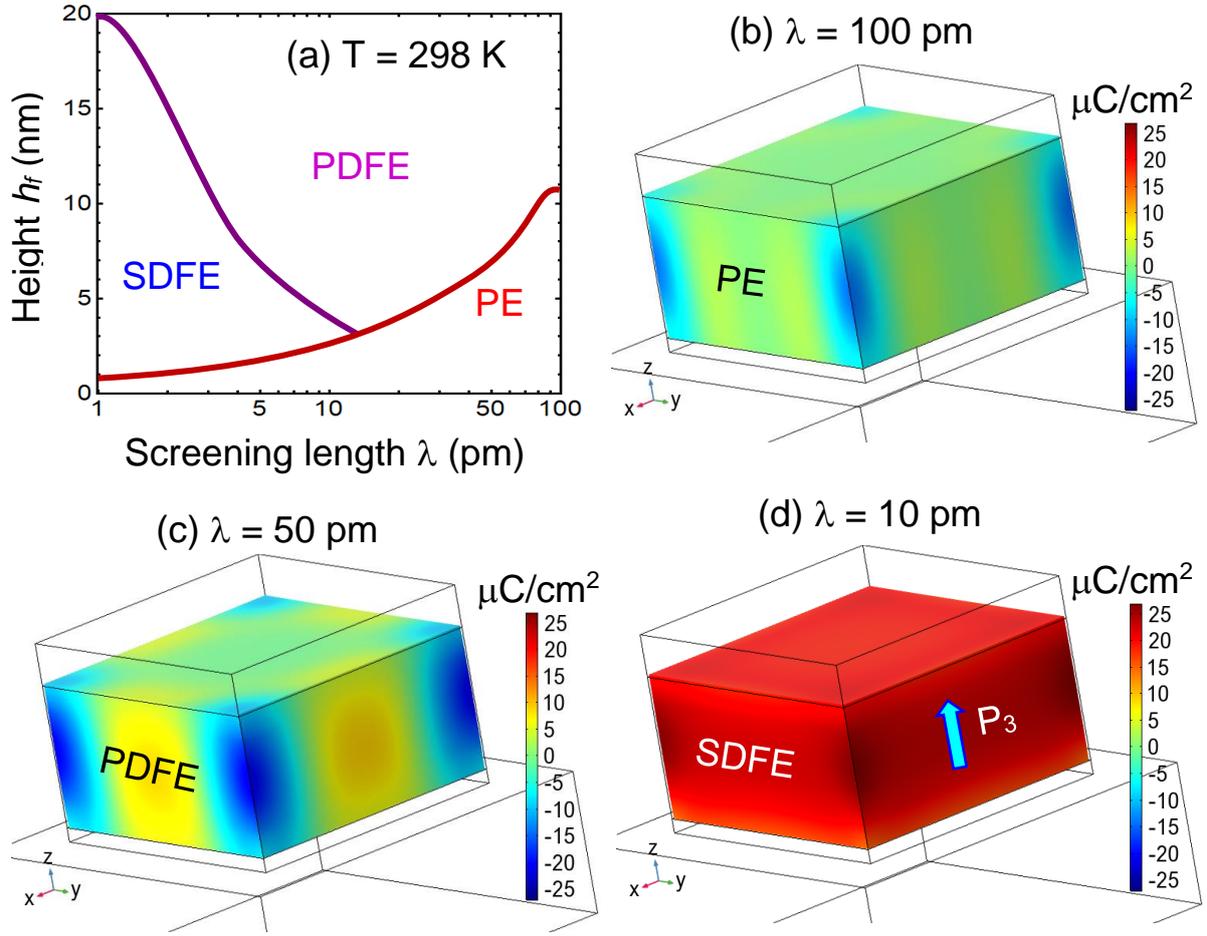

**FIGURE 2**. **(a)** Phase diagram of the SPS core of thickness $h_f$ sandwiched between two dielectric layers of thickness $d_1$ and $d_2$. **(b)-(d)** The equilibrium polarization distribution in the SPS nanoparticle calculated for $h_f = 4$ nm and several values of the effective screening length $\lambda = 100$ pm **(b)**, 50 pm **(c)**, and 10 pm **(d)**. The direction of the uniaxial polarization $P_3$ is shown by the blue arrow in **(d)**. Other parameters: $w = 8$ nm, $d_1 = 0.4$ nm, $d_2 = 1.2$ nm, $\varepsilon_l = 100$, $\varepsilon_u = 10$, $u_m = +1\%$, and $T = 298$ K.

The FEM results show that the screening charges at the top and bottom surfaces of the ferroelectric core, $z = d_1$ and $z = h - d_2$, can influence on the phase transitions and polar properties of the nanoparticle core. Note that the ferroelectric polarization cannot rotate in uniaxial ferroelectrics. Hence, the polarization directed along z-axis is parallel to the sidewalls and it does not require any screening at the walls.

Below we will use the ranges of nanoparticle sizes and surface charge parameters for which the domain formation does not take place in the core. In a single domain state, the Landau-Khalatnikov equation determining the average polarization $\bar{P}$ has the form [21]:

$$\Gamma \frac{d}{dt}\bar{P} + \tilde{\alpha}(T)\bar{P} + \tilde{\beta}\bar{P}^3 + \gamma\bar{P}^5 = \frac{\Psi}{h}. \qquad (6a)$$

Here the coefficients $\tilde{\alpha}(T) = \alpha_T(T - T_C) - 2\frac{Q_{13}+Q_{23}}{s_{11}+s_{12}}u_m$ and $\tilde{\beta} = \beta + \frac{(Q_{13}+Q_{23})^2}{s_{11}+s_{12}} + \frac{(Q_{13}-Q_{23})^2}{s_{11}-s_{12}}$ are renormalized by elastic strains originated from the nanoparticles-substrate lattice mismatch [61, 62]. Using the results of Ref.[23] and **Appendix A**, the overpotential $\Psi$ can be determined in a self-consistent manner:

$$\frac{\Psi}{h} = \frac{\delta\,\sigma}{\varepsilon_0\varepsilon_b(\eta+\delta)} - \frac{\delta\bar{P}}{\varepsilon_0\varepsilon_b(\eta+\delta)} + \frac{V(t)}{\varepsilon_b(\eta+\delta)}, \qquad (6b)$$

where the effective thickness of dielectric layers is defined as $\delta = \frac{d_1}{\varepsilon_l} + \frac{d_2}{\varepsilon_u}$, the reduced core thickness is defined as $\eta = \frac{h_f}{\varepsilon_b}$, and the total surface charge density $\sigma = \sigma_p + \sigma_n$. From Eq.(6b), $\Psi$ contains a contribution which is proportional to the surface charge density $\sigma$, a depolarization field contribution which is proportional to $\bar{P}$, and an external potential drop which is proportional to the applied voltage $V(t)$. The surface charge and the depolarization field contributions, $\frac{\delta\,\sigma}{\varepsilon_0\varepsilon_b(\eta+\delta)}$ and $\frac{\delta\bar{P}}{\varepsilon_0\varepsilon_b(\eta+\delta)}$, do not depend on the particle lateral size, $w$, because the polarization cannot rotate in uniaxial ferroelectrics; therefore, the screening charges at the sidewalls cannot influence on the polar properties and phase transitions of the nanoparticle core. Below we consider the case of a periodic applied voltage, $V(t) = V \cdot \sin(\omega t)$, where $V$ is the amplitude and $\omega = \frac{2\pi}{\tau_V}$ is the frequency.

Using Eqs.(1)-(2), the nonlinear dynamics of the positive and negative surface charges obeys relaxation equations:

$$\tau_p \frac{\partial \sigma_p}{\partial t} + \sigma_p = eZ_p C_p \left(1 + g_p \exp\left(\frac{\Delta G_p^0 + eZ_p\Psi}{k_B T}\right)\right)^{-1}, \qquad (6c)$$

$$\tau_n \frac{\partial \sigma_n}{\partial t} + \sigma_n = eZ_n C_n \left(1 + g_n \exp\left(\frac{\Delta G_n^0 + eZ_n\Psi}{k_B T}\right)\right)^{-1}. \qquad (6d)$$

Hence, we obtained a coupled system of four nonlinear coupled equations, Eqs.(4), with three different characteristic time-scales, $\Gamma$, $\tau_p$, and $\tau_n$. These times can be very different from each other for the case of ionic-electronic screening, as well as different from the period $\tau_V$ of applied voltage $V(t)$ in Eq.(6b). Since a polarization relaxation is determined by soft optical phonons, the strong inequality, $\Gamma/|\alpha| \ll \tau_{n,p}$, is valid far from the Curie temperature; and it makes sense to normalize time $t$ in Eqs.(6) to the Khalatnikov time, $\tau_{Kh} = \Gamma/|\alpha_T T_c|$. The normalization is used hereinafter.

### IV. POLARIZATION AND SUSCEPTIBILITY UNDER THE PRESENCE OF A NONLINEAR SURFACE SCREENING

Since the system of the coupled Eqs.(6) does not allow for analytical solutions, we studied the numerical solutions for polarization, screening charges, and dielectric susceptibility voltage dependences in the actual range of the nanoparticle sizes $h_f \sim (2-20)$ nm, dielectric layer thicknesses $d_{1,2} \sim (0-2)$ nm, temperatures $\sim (100-350)$ K, concentrations $C_{n,p} \sim (10^{16} - 10^{18})$ m$^{-2}$, formation energies $\Delta G_{n,p}^0 \sim (0.03-0.3)$ eV [18, 19, 20] of the screening charge, and their relaxation times $1 \leq \tau_{n,p} \leq 10^3$ (in the units of $\tau_{Kh}$). The frequency $\omega$ of the applied voltage $V(t)$ is very low, $\omega \tau_{Kh} \ll 1$, and so its period $\tau_V \gg 1$. As uniaxial ferroelectric we use the SPS core, which parameters are listed in **Table SI** in Supplement. A tensile mismatch strain $u_m = +1\%$ is applied to support the spontaneous polarization in the ferroelectric core. Numerical results were visualized in Mathematica 12.2 [63].

It was found that the behavior of polarization and susceptibility is very sensitive to the concentrations, formation energies, and relaxation times of the screening charges. In general, by increasing the concentrations $C_{n,p}$ one can switch the core state between the PE-like, AFEI, mixed AFEI-FEI, FEI, and FE-like FEI states. The typical features of these transitions are shown schematically in **Fig. 3(a)-(b)**. At that the dependence of the average polarization on applied voltage, $\bar{P}(V)$, is symmetric for equal relaxation times $\tau_n = \tau_p$ [**Fig. 3(a)**] and becomes strongly asymmetric for $\tau_n \ll \tau_p$ [**Fig. 3(b)**].

From **Fig. 3(a)** the continuous transition between the PE-like, AFEI, AFE-FEI, and FE-like FEI states takes place when the surface charges concentration $C_n = C_p$ increases from $10^{16}$ m$^{-2}$ to $10^{18}$ m$^{-2}$ and $\tau_n = \tau_p$. The voltage dependences of the polarization $\bar{P}(V)$, negative and positive surface charges, $\sigma_n(V)$ and $\sigma_p(V)$, and their difference - the "screening efficiency", $\bar{P}(V) - \sigma(V)$, are quasilinear and hysteresis-less in the PE-like state, which exists at concentrations $C_{n,p} \leq 10^{16}$ m$^{-2}$. When the concentrations increase up to $5 \cdot 10^{16}$ m$^{-2}$, an antiferroelectric-type double hysteresis loop $\bar{P}(V)$ appears. The voltage positions of the polarization double loops opening almost coincide with the positions of the "minor" loops openings of $\sigma_n(V)$ (at negative voltages) and $\sigma_p(V)$ (at positive voltages) which are well-separated. The minor loops $\sigma_n(V)$ and $\sigma_p(V)$ open at some critical voltages $V_{cr}$ due to the strongly nonlinear exponential dependence of $\sigma_{n,p}(\varphi)$ given by Eqs.(2). The behavior corresponds to the AFEI state induced in the ferroelectric core by the interaction between the ferroelectric dipoles and the surface screening charges at $V > |V_{cr}|$. Further increase of the screening charge concentration above $5 \cdot 10^{16}$ m$^{-2}$ leads to the appearance of the pinched $\bar{P}(V)$ loops and the overlapping of the $\sigma_n(V)$ and $\sigma_p(V)$ minor loops. The behavior corresponds to the mixed AFEI-FEI state in the ferroelectric core. The FEI state of the core, which is characterized by a single FE-type hysteresis loop $\bar{P}(V)$, is induced by the screening charges with the concentration above $10^{17}$ m$^{-2}$. The loops in the FEI state become seemingly indistinguishable from the FE loops at $C_{n,p} > 5 \cdot 10^{17}$ m$^{-2}$. However, this is an apparent effect only, because it is the FE-like FEI state supported by the nonlinear dynamics of surface charge, and the state does not exist without the nonlinear screening. The statement is grounded by the complex view of $\sigma_n(V)$ and $\sigma_p(V)$ loops with straight lines, sharp edges, and very sharp maxima at the coercive voltage. These unusual peculiarities only occur at high concentrations $C_{n,p} > 10^{18}$ m$^{-2}$. This suggests that in realistic materials other charge relaxation mechanisms can become active (desorption, charge emission), rendering these states unrealizable.

From **Fig. 3(b)** the continuous transition between the PE-like and strongly asymmetric FEI states takes place when the surface charges concentration $C_n = C_p$ increases from $10^{16}$ m$^{-2}$ to $10^{18}$ m$^{-2}$ and $\tau_n \ll \tau_p$. The asymmetry originates from the strong retardation of the screening by one type of the carriers with respect to the other, and with respect the period $\tau_V$ of the applied voltage ($\omega \tau_p \gg 1$). The retardation of dynamical screening is responsible for both the asymmetry and the significant horizontal shift of the polarization and surface charge loops, as well as for the disappearance of the double AFEI loops. Note that the case $\tau_n \gg \tau_p$ leads to the same physical picture as $\tau_n \ll \tau_p$, with the substitution $V \to -V$. In particular, left-shifted loops at $\tau_n \ll \tau_p$ becomes right-shifted at $\tau_n \gg \tau_p$, because the dynamic properties of positive and negative screening charges interchange.

**Figures 3(b)-(d)** illustrates schematically the dependence of the nanoparticle polar state on the formation energies $\Delta G_{n,p}^0$ of the surface charges. The simultaneous decrease of both formation energies from 0.15eV to 0.01eV leads to the continuous transition from the AFEI to the FEI state in the ferroelectric core [**Fig. 3(c)**]. The decrease of one of the formation energies, $\Delta G_n^0$, from 0.15eV to 0.01eV with the other fixed, $\Delta G_p^0 = 0.15$eV, leads to the mixing of the AFEI and FEI states [**Fig. 3(d)**]. The hysteresis loops in the mixed AFEI-FEI state are typically strongly asymmetric, shifted horizontally, distorted, and can be significantly pinched. Since the antiferroelectric-like double loops and pinched loops of polarization are often observed in polydomain ferroelectric thin films [14], as well as the fact that they are typically associated with polydomain or vortex-like domain states in ferroelectric nanoparticles [3, 10], their appearance in a single-domain ferroelectric core covered with ionic-electronic screening charges seems unusual and requires further studies. In the considered case, the appearance of the antiferroelectric-like loops is caused by the minor loops of the surface charges, $\sigma_n(V)$ and $\sigma_p(V)$, which are absent at small voltages and open at $|V| > V_{cr}$; the critical voltage $V_{cr}$ depends on the nanoparticles thickness and temperature [see the blue and red loops in **Figs. 3(c)** and **3(d)**]. Note, the convergence of the charge loops to the point of full overlap as we step through the different phases, e.g., from PE-like

to FE-like FEI phases [see **Figs. 3(a)** and **3(c)**]. This corroborates the surface charge origin of the FE-like hysteresis loops of polarization in the different phases.

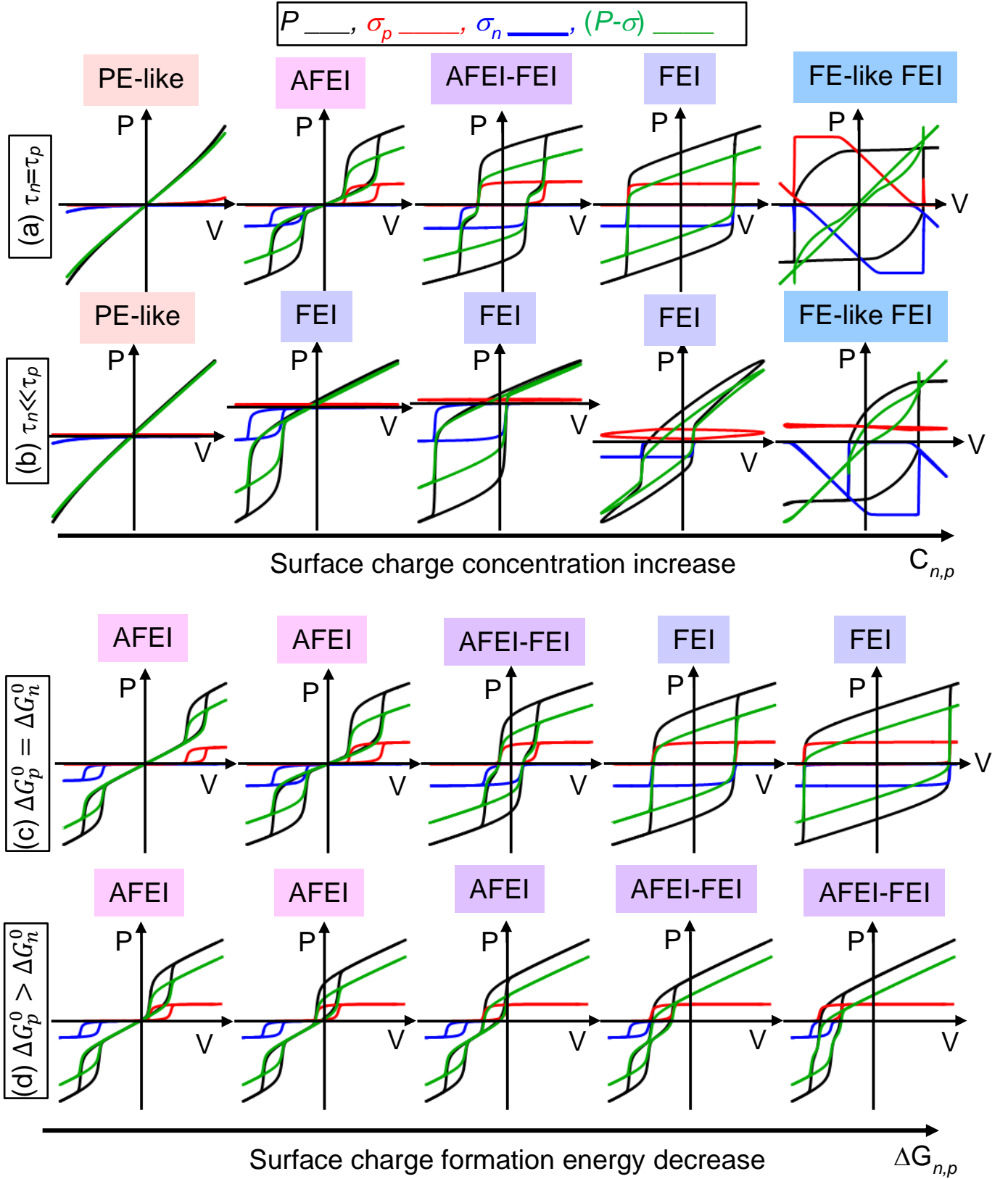

**FIGURE 3.** The quasi-static voltage dependences of the polarization $\bar{P}(V)$ (black curves), positive (red curves) and negative (blue curves) surface charges, $\sigma_p(V)$ and $\sigma_n(V)$, and the screening efficiency $\bar{P}(V) -$

$\sigma(V)$ (green curves) calculated for different concentrations, formation energies, and relaxation times of the screening charges. The surface charges concentrations $C_n = C_p$ increase from $10^{16}$ m$^{-2}$ to $10^{18}$ m$^{-2}$ for **(a)** and **(b)**, which are calculated for $\Delta G_p^0 = \Delta G_n^0 = 0.1$ eV. The relaxation times $\tau_n = \tau_p = 10\tau_{Kh}$ for **(a)**, and $\tau_n = 10\tau_{Kh}$, $\tau_p \gg 10^3 \tau_{Kh}$ for **(b)**. The surface charges formation energies $\Delta G_p^0 = \Delta G_n^0$ decreases from 0.15 eV to 0.01 eV for the part **(c)**. Part **(d)** corresponds to $\Delta G_p^0 = 0.15$ eV and $\Delta G_n^0$ decreasing from 0.15 eV to 0.01 eV. For parts **(c)** and **(d)** $C_n = C_p = 5 \cdot 10^{16}$ m$^{-2}$ and $\tau_n = \tau_p = 10\tau_{Kh}$. For all plots $h_f = 4$ nm, $w = 8$ nm, $d_1 = 0.4$ nm, $d_2 = 1.2$ nm, $\varepsilon_l = 100$, $\varepsilon_u = 10$, $u_m = +1\%$, and $T = 298$ K.

Note that for both cases, $\tau_n = \tau_p$ and $\tau_n \ll \tau_p$, where $C_{n,p} < 10^{18}$ m$^{-2}$, the polarization, $\bar{P}(V)$, and the screening efficiency, $\bar{P}(V) - \sigma(V)$, follow the same trends, meaning that the total screening charge $\sigma(V)$ correlates with the polarization changes in order to partially screen it. At the same time the voltage behavior of $\sigma_n(V)$ and $\sigma_p(V)$ are complementary to each other at $\tau_n = \tau_p$, as they can screen the bound charge of the opposite polarity [compare the blue, red, and green curves in **Figs. 3(a), (c)-(d)**]. Let us underline that the difference $\bar{P}(V) - \sigma(V)$ can be measured experimentally as the charge on = one of the electrodes.

To explore the physical origin and quantify the trends shown in **Fig. 3**, we consider the case $\tau_n = \tau_p$, corresponding to the surface screening by electrons and holes; and the case $\tau_n \ll \tau_p$ corresponding to the surface screening by electrons and cations, respectively. Results are presented in two subsections below; figures in these subsections illustrate the typical voltage dependences of the average polarization $\bar{P}(V)$, negative and positive surface charges, $\sigma_n(V)$ and $\sigma_p(V)$, and the difference $\bar{P}(V) - \sigma(V)$. We also show the derivatives, $\frac{\partial \bar{P}}{\partial V}$ and $\frac{\partial}{\partial V}(\bar{P} - \sigma)$, which are directly related with the dielectric susceptibility and "effective" capacitance, respectively. In all figures the frequency $\omega$ of the applied voltage is very low, $\omega \tau_{Kh} \leq 10^{-3}$; here we only show the stationary loops and omit transient processes. The third subsection explores the role of size effects on the polarization and susceptibility hysteresis loops.

### A. The influence of surface screening electrons and holes on the polar state and dielectric susceptibility

**Figures 4-7** illustrate that the transition between symmetric PE-like, AFEI, FEI, and FE-like states occurs under the increase of the surface charges concentration $C_n = C_p$ from $2 \cdot 10^{16}$ m$^{-2}$ to $10^{18}$ m$^{-2}$. These dependences are calculated for a 4-nm SPS core at room temperature and several values of equal relaxation times $\tau_n = \tau_p = 10\tau_{Kh}$, $\tau_n = \tau_p = 10^2 \tau_{Kh}$, and $\tau_n = \tau_p = 10^3 \tau_{Kh}$ corresponding to the surface screening by electrons and holes. The top rows in **Figs. 4-7** show

voltage dependences of the polarization $\bar{P}(V)$, positive and negatives surface charges, $\sigma_p(V)$ and $\sigma_n(V)$, and the difference $\bar{P}(V) - \sigma(V)$. The bottom rows show the quasi-static dependences of the dielectric susceptibility $\frac{\partial \bar{P}}{\partial V}$ and effective capacitance $\frac{\partial}{\partial V}(\bar{P} - \sigma)$, which can have four or two symmetric maxima indicating on the AFEI or the FEI states, respectively (see **Figs. 4-6**). The voltage dependences of the susceptibility are much more complex in FE-like FEI states (see **Fig. 7**). The common peculiarities of the figures are the smearing of the loop details, the significant decrease of the polarization saturation rate as $\tau_{n,p}$ increases, and the strong and nonlinear increase of the coercive field (i.e., loop width) as $C_{n,p}$ and $\tau_{n,p}$ increase. To show the saturated loops whenever it is possible, we use different voltage scales in **Figs. 4-7**. Also, there are several important differences in **Figs. 4-7**, which are discussed below.

In **Fig. 4** the magnitude of polarization and the coercive voltage, corresponding to the appearance of very thin AFEI-type loops, do not vary significantly between the plots **(a)** and **(b)**, regardless of the 10 times increase in $\tau_{n,p}$ (from 10 to $10^2$). The opening of the double loops becomes evident from both the loop shape in **Fig. 4(b)** and especially from the four well-separated susceptibility maxima in **Fig. 4(e)**, while at 10 times longer relaxation times, $\tau_{n,p} = 10^3$, a very slim polarization loop with a small remanent polarization appears [**Fig. 4(c)**]. The loop has at least a four times higher maximal polarization corresponding to the six times higher voltage amplitude in comparison with the loops shown in **Fig. 4(a)-(b)**. The unsaturated loops, similar to the loop in **Fig. 4(c)**, are typical for ferroelectric relaxors; and a relaxor-like character is seen from the susceptibility plot **(f)**, where the maxima at coercive voltages are very diffuse and shifted towards the central point $V = 0$.

As one can see from **Figs. 5(a)-(c)**, an increase of the concentrations $C_{n,p}$ by a factor of 2.5 [in comparison with **Figs. 4(a)-(c)**] leads to an opening of double AFEI-type polarization loops at small relaxation times ($\tau_{n,p} = 10$), a pinching of single FEI-type polarization loops at longer relaxation times ($\tau_{n,p} = 10^2$), and a relatively moderate blowing of the relaxor-like loops at very long relaxation times ($\tau_{n,p} = 10^3$). These trends are evident from the comparison of the susceptibility maxima position, shape, and height in **Figs. 5(d)-(f)** with **Figs. 4(d)-(f)**. Actually, we see four very sharp and well-separated pairwise maxima in **Fig. 5(d)**; these maxima become much wider with a reduced height when $\tau_{n,p}$ is increased by factor of ten [see **Fig. 5(e)**]. The voltage separation of the pairwise maxima almost disappears at very long values of $\tau_{n,p}$ [see **Fig. 5(f)**]. Note that **Fig. 5** has different voltage scales, which are ±0.04 V for the plots **(a, d)**, ±0.1 V for the plots **(b, e)**, and ±0.6 V for the plots **(c, f)**.

Note that the doubling of the concentrations $C_{n,p}$ leads to a significant increase in the remanent polarization, its maximal value, and the coercive voltages at $\tau_{n,p} = 10^2$ [compare **Fig. 5(b)** with **4(b)**]. However, the relaxor-like loops, observed at $\tau_{n,p} = 10^3$, are almost insensitive to the concentration increase [compare **Fig. 5(c)** with **4(c)**]. This can be explained by the very strong retardation of the polarization screening with respect to the polarization changes taking place over very long relaxation times. We believe that the retardation effect leads to the relaxor-like loops shown in **Figs. 4(c)** and **5(c)**.

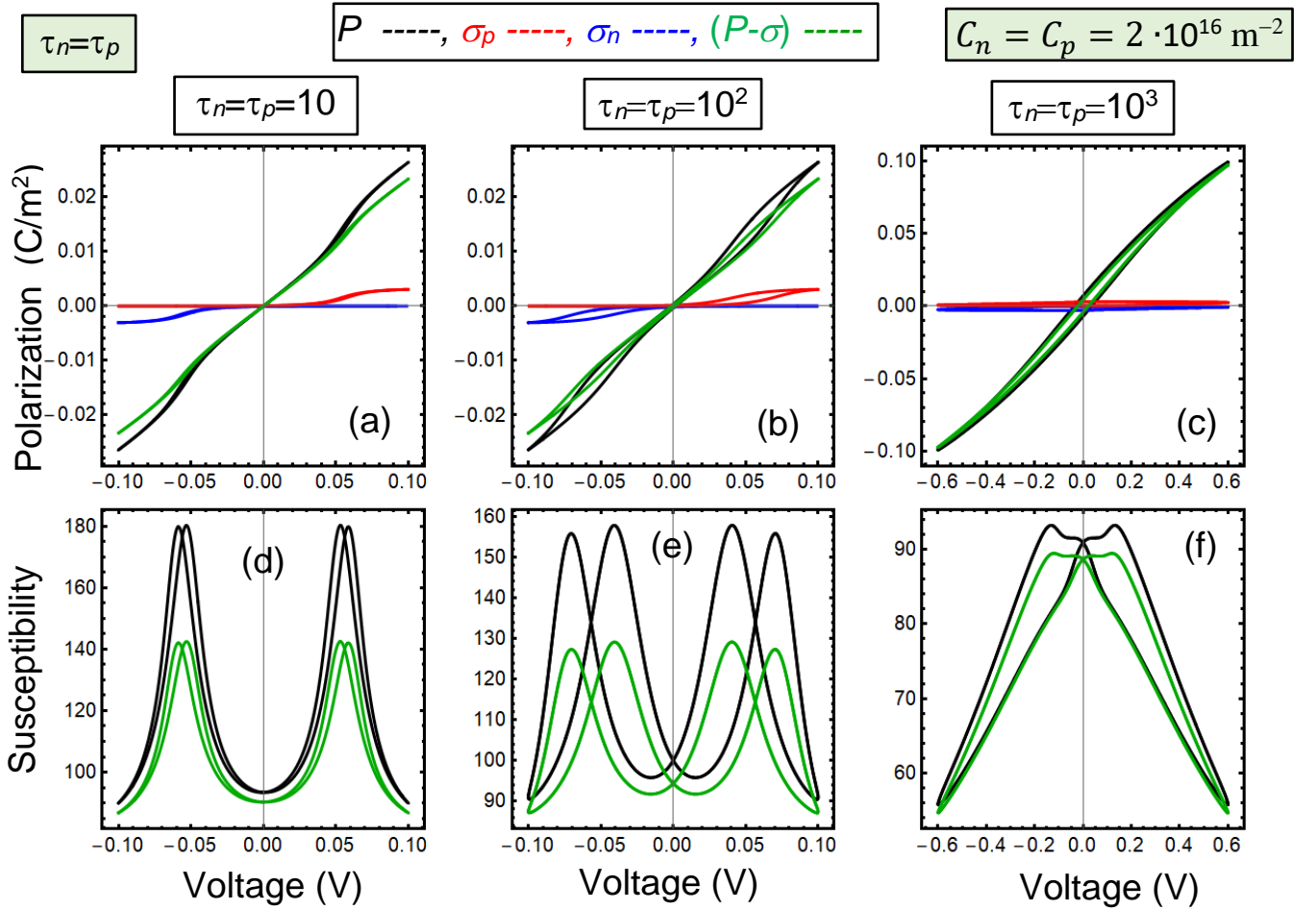

**FIGURE 4. Symmetric paraelectric-like antiferroionic and ferroionic states.** (**a, b, c**) The quasi-static voltage dependences of the polarization $\bar{P}(V)$ (black curves), positive $\sigma_p(V)$ (red curves) and negative $\sigma_n(V)$ (blue curves) surface charges, and the difference $\bar{P}(V) - \sigma(V)$ (green curves). (**d, e, f**) The quasi-static dependences of the dielectric susceptibility $\frac{\partial \bar{P}}{\partial V}$ (black curves) and effective capacitance $\frac{\partial}{\partial V}(\bar{P} - \sigma)$ (green curves). The dependences are calculated for the surface charge concentrations $C_n = C_p = 2 \cdot 10^{16}$ m$^{-2}$ and several values of relaxation times $\tau_n = \tau_p = 10\tau_{Kh}$ (**a, d**), $\tau_n = \tau_p = 10^2\tau_{Kh}$ (**b, e**), and $\tau_n = \tau_p = 10^3\tau_{Kh}$ (**c, f**). Other parameters: $h_f = 4$ nm, $w = 8$ nm, $d_1 = 0.4$ nm, $d_2 = 1.2$ nm, $\varepsilon_l = 100$, $\varepsilon_u = 10$, $u_m = +1\%$, and $T = 298$ K, $Z_p = -Z_n = +1$, $g_p = g_n = 1$, and $\Delta G_p^0 = \Delta G_n^0 = 0.1$ eV.

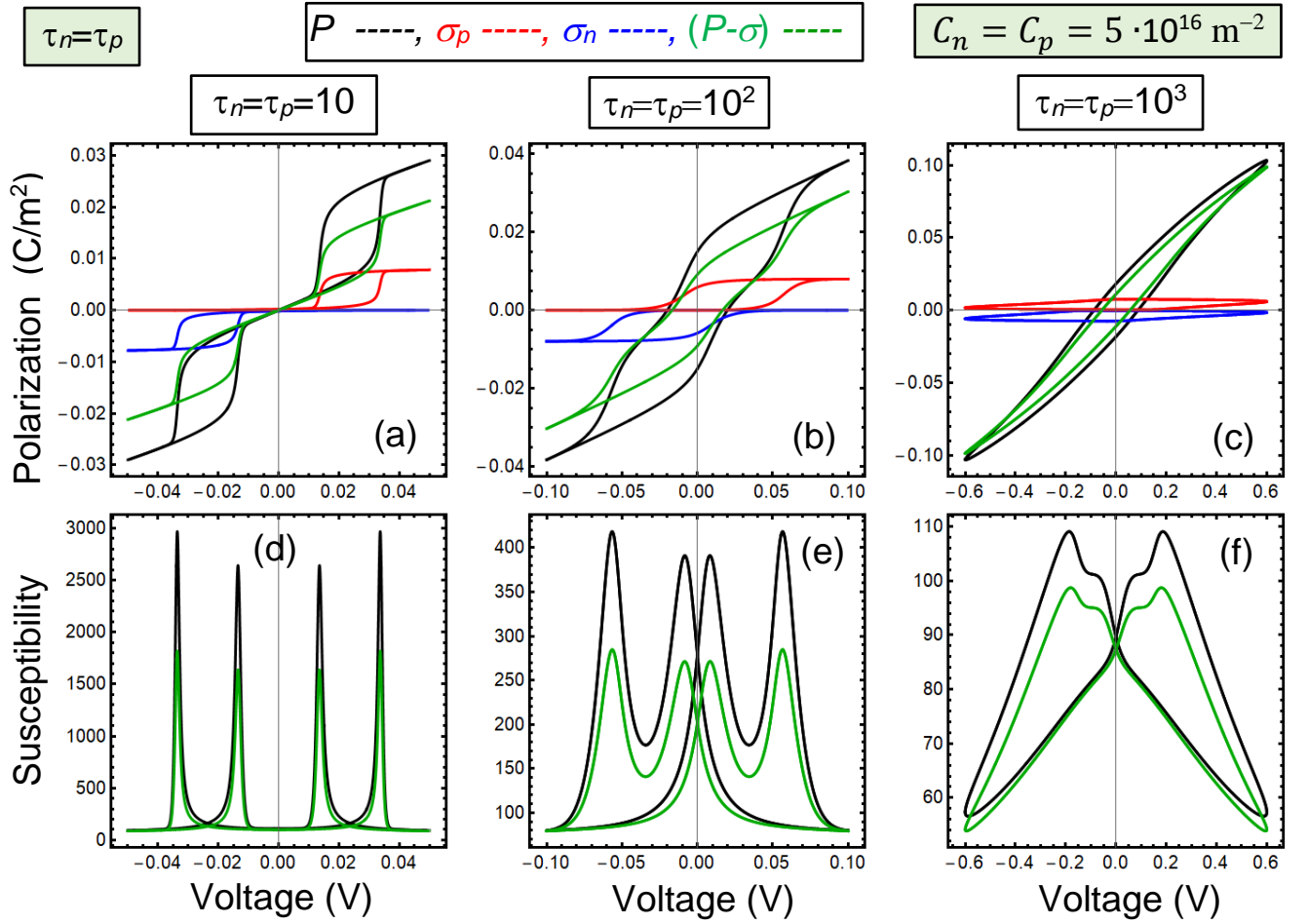

**FIGURE 5. Symmetric antiferroinic and ferroionic states. (a, b, c)** The quasi-static voltage dependences of the polarization $\bar{P}(V)$ (black curves), positive $\sigma_p(V)$ (red curves) and negative $\sigma_n(V)$ (blue curves) surface charges, and the difference $\bar{P}(V) - \sigma(V)$ (green curves). **(d, e, f)** The quasi-static dependences of the dielectric susceptibility $\frac{\partial \bar{P}}{\partial V}$ (black curves) and effective capacitance $\frac{\partial}{\partial V}(\bar{P} - \sigma)$ (green curves). The dependences are calculated for the surface charge concentrations $C_n = C_p = 5 \cdot 10^{16}$ m$^{-2}$ and several values of relaxation times $\tau_n = \tau_p = 10$ **(a, d)**, $\tau_n = \tau_p = 10^2$ **(b, e)**, and $\tau_n = \tau_p = 10^3$ **(c, f)** in the units of $\tau_{Kh}$. Other parameters are the same as in **Fig. 4**.

As it can be seen from **Figs. 6(a)-(c)**, a further increase of the concentrations $C_{n,p}$ by a factor of two [in comparison with **Figs. 5(a)-(c)**] leads to an opening of single FEI-type polarization loops at smaller ($\tau_{n,p} = 10$) and longer ($\tau_{n,p} = 10^2$) relaxation times, and more pronounced saturated relaxor-like loops at very long relaxation times ($\tau_{n,p} = 10^3$). These trends are confirmed by the dielectric susceptibility features, shown in **Figs. 6(d)-(f)**. Opposed to the AFEI state shown in **Fig. 5(d)**, there are only two very sharp and well-separated maxima in **Fig. 6(d)**. The main maxima become much wider and lower with increasing $\tau_{n,p}$, while small secondary maxima appear [see

**Fig. 6(e)**]. The separation of the main and secondary maxima disappears for very long relaxation times in **Fig. 6(f)**. Note that different parts in **Figs. 6** have ten times different voltage scales, which are ±0.2 V for **(a - d)** and ±2 V for **(c, f)**.

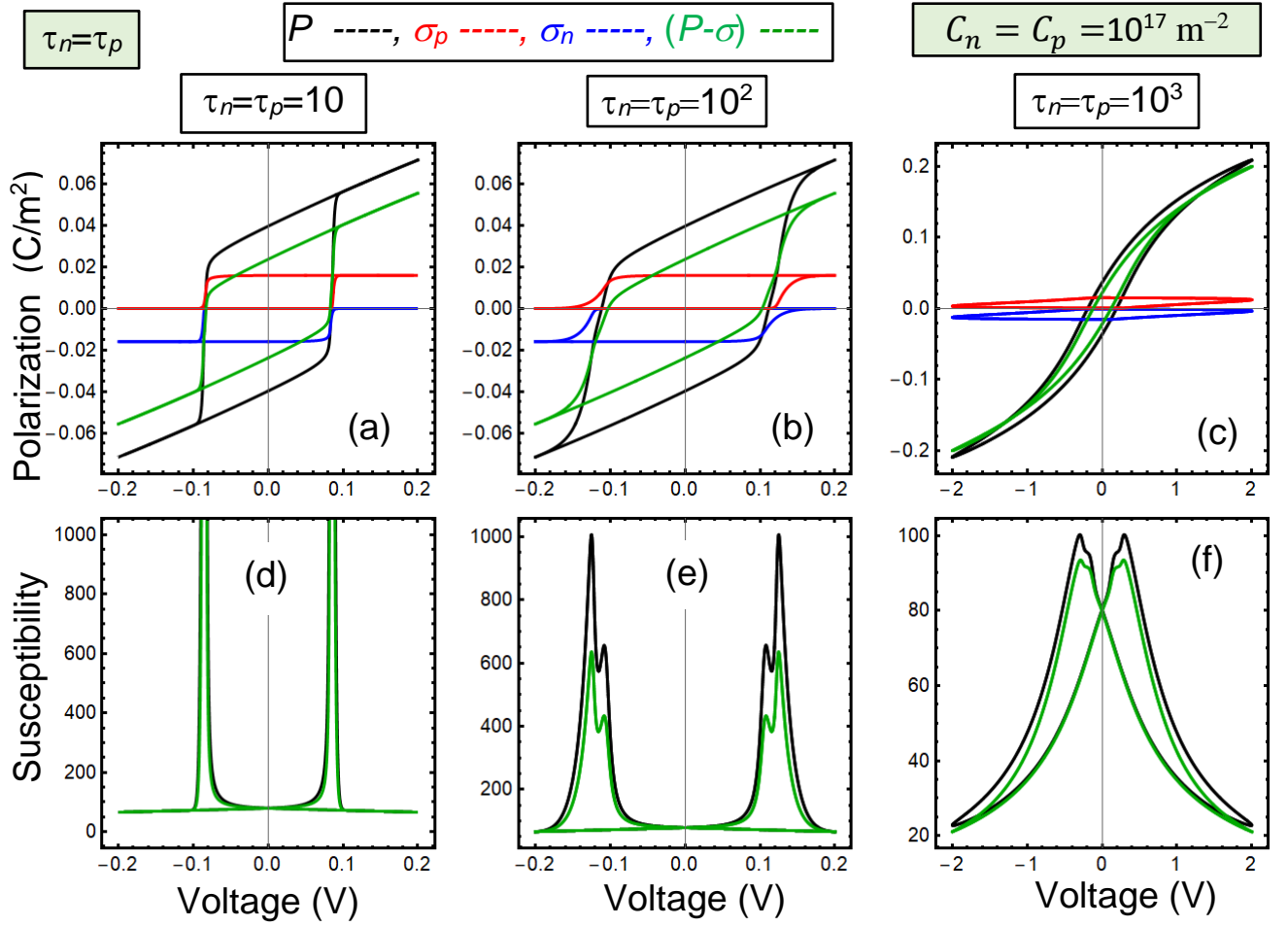

**FIGURE 6. Symmetric ferroionic states. (a, b, c)** The quasi-static voltage dependences of the polarization $\bar{P}(V)$ (black curves), positive $\sigma_p(V)$ (red curves) and negative $\sigma_n(V)$ (blue curves) surface charges, and the difference $\bar{P}(V) - \sigma(V)$ (green curves). **(d, e, f)** The quasi-static dependences of the dielectric susceptibility $\frac{\partial \bar{P}}{\partial V}$ (black curves) and effective capacitance $\frac{\partial}{\partial V}(\bar{P} - \sigma)$ (green curves). The dependences are calculated for the surface charge concentrations $C_n = C_p = 10^{17}$ m$^{-2}$ and several values of relaxation times $\tau_n = \tau_p = 10$ **(a, d)**, $\tau_n = \tau_p = 10^2$ **(b, e)**, and $\tau_n = \tau_p = 10^3$ **(c, f)** in the units of $\tau_{Kh}$. Other parameters are the same as in **Fig. 4**.

As it can be seen from **Fig. 7(a)-(b)**, the increase of the concentrations $C_{n,p}$ by a factor of 10 [in comparison with **Fig. 6**] at smaller ($\tau_{n,p} = 10$) and longer ($\tau_{n,p} = 10^2$) relaxation times leads to the appearance of well-saturated FE-like square-shaped polarization loops with a high coercive voltage (~2 V). At very long relaxation times ($\tau_{n,p} = 10^3$) the loops are less saturated and more smooth-shaped [see **Fig. 7(c)**]. The complete saturation of the polarization loops taking place at the voltages $|V| > 1$ V leads to unusual properties of the dielectric susceptibility, shown in **Figs. 7(d)-(f)**. Here, besides two very sharp and well-separated maxima of $\frac{\partial \bar{P}}{\partial V}$, we also have wide regions of nearly zero susceptibility, $\frac{\partial \bar{P}}{\partial V} \approx 0$, and very narrow regions of a negative effective capacitance $\frac{\partial}{\partial V}(\bar{P} - \sigma)$. Both of these peculiarities, $\frac{\partial \bar{P}}{\partial V} \approx 0$ and $\frac{\partial}{\partial V}(\bar{P} - \sigma) < 0$, can be treated as signatures of the negative capacitance of the core, which can be quasi-steady state since the external field frequency is very low. Note that **Fig. 7** has slightly different voltage scales, which are ±3 V for **(a - d)** and ±4 V for **(c, f)**.

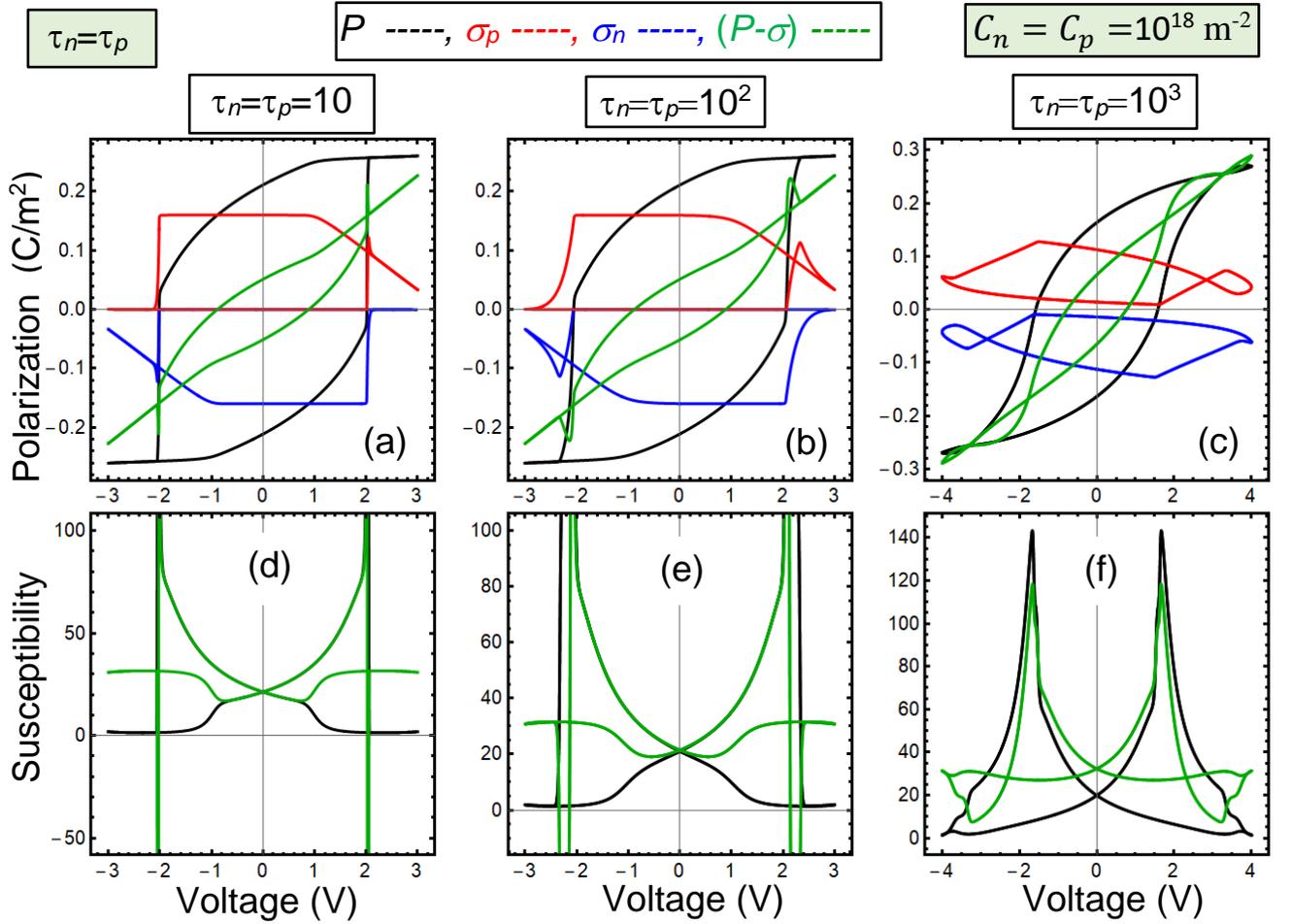

**FIGURE 7. Symmetric ferroelectric-like ferroionic states. (a, b, c)** The quasi-static voltage dependences of the polarization $\bar{P}(V)$ (black curves), positive $\sigma_p(V)$ (red curves) and negative $\sigma_n(V)$ (blue curves) surface charges, and the difference $\bar{P}(V) - \sigma(V)$ (green curves). **(d, e, f)** The quasi-static dependences of the dielectric susceptibility $\frac{\partial \bar{P}}{\partial V}$ (black curves) and effective capacitance $\frac{\partial}{\partial V}(\bar{P} - \sigma)$ (green curves). The dependences are calculated for the surface charge concentrations $C_n = C_p = C = 10^{18}$ m$^{-2}$ and several

values of relaxation times $\tau_n = \tau_p = 10$ (**a, d**), $\tau_n = \tau_p = 10^2$ (**b, e**), and $\tau_n = \tau_p = 10^3$ (**c, f**) in the units of $\tau_{Kh}$. Other parameters are the same as in **Fig. 4**.

## B. The influence of surface screening by electrons and cations on the polarization state and susceptibility

**Figures 8-11** illustrate that the transition between asymmetric PE-like, AFEI, FEI, and almost symmetric FE-like states occurs with an increase of the surface charges concentrations $C_{n,p}$ from $2 \cdot 10^{16}$ m$^{-2}$ to $10^{18}$ m$^{-2}$ under different relaxation times $\tau_n \ll \tau_p$, corresponding to the surface screening by electrons and cations. These dependences are calculated for a 4-nm SPS core at room temperature, an electron relaxation time $\tau_n = 1\tau_{Kh}$, and several values of much higher cation relaxation times $\tau_p = 10\tau_{Kh}$, $\tau_p = 10^2\tau_{Kh}$, and $\tau_p = 10^3\tau_{Kh}$. Note that the coercive field (i.e., loop width) increases strongly and nonlinearly with $C_{n,p}$.

The top rows show quasi-static voltage dependences of the polarization $\bar{P}(V)$, positive and negative surface charges, $\sigma_p(V)$ and $\sigma_n(V)$, and the difference $\bar{P}(V) - \sigma(V)$. The bottom rows show the quasi-static dependences of the dielectric susceptibility $\frac{\partial \bar{P}}{\partial V}$ and effective capacitance $\frac{\partial}{\partial V}(\bar{P} - \sigma)$, which can have two or four asymmetric maxima indicating AFEI, AFEI-FEI, or FEI states, respectively (see **Figs. 8-10**). Furthermore, the concentration of surface charges $C_{n,p} \cong 10^{17}$m$^{-2}$ gives more typical polarization and susceptibility hysteresis loops [shown in **Fig. 10**] compared to the other cases. For $C_{n,p} \geq 10^{18}$m$^{-2}$, the voltage dependences of the surface charge density and susceptibility are much more complex and unusual, they correspond to the FE-like FEI states shown in **Fig. 11**.

The common peculiarity of all figures is the smearing of the asymmetric loops with an increasing $\tau_p$. The asymmetry of the left and right parts of the loops are related to the fact that the electron retardation is much smaller than it is for cations. Since the cation density increases and saturates at positive overpotentials [see **Fig. 1(c)**], most of the loops in **Figs. 8-9** are left-shifted and their openings are bigger at positive voltages.

Other comments and analyses made in the previous subsection regarding the symmetric polarization loops and susceptibility maxima shown in **Figs. 4-7** remain relevant for their asymmetric and/or shifted voltage dependences shown in **Figs. 8-11**. Note that the asymmetry of the polarization and susceptibility voltage dependences becomes much weaker with a $C_{n,p}$ increase above $10^{17}$ m$^{-2}$; therefore, **Figs. 10** and **11** looks similar to **Figs. 6** and **7** except for the horizontal shift of the polarization loops and susceptibility maxima.

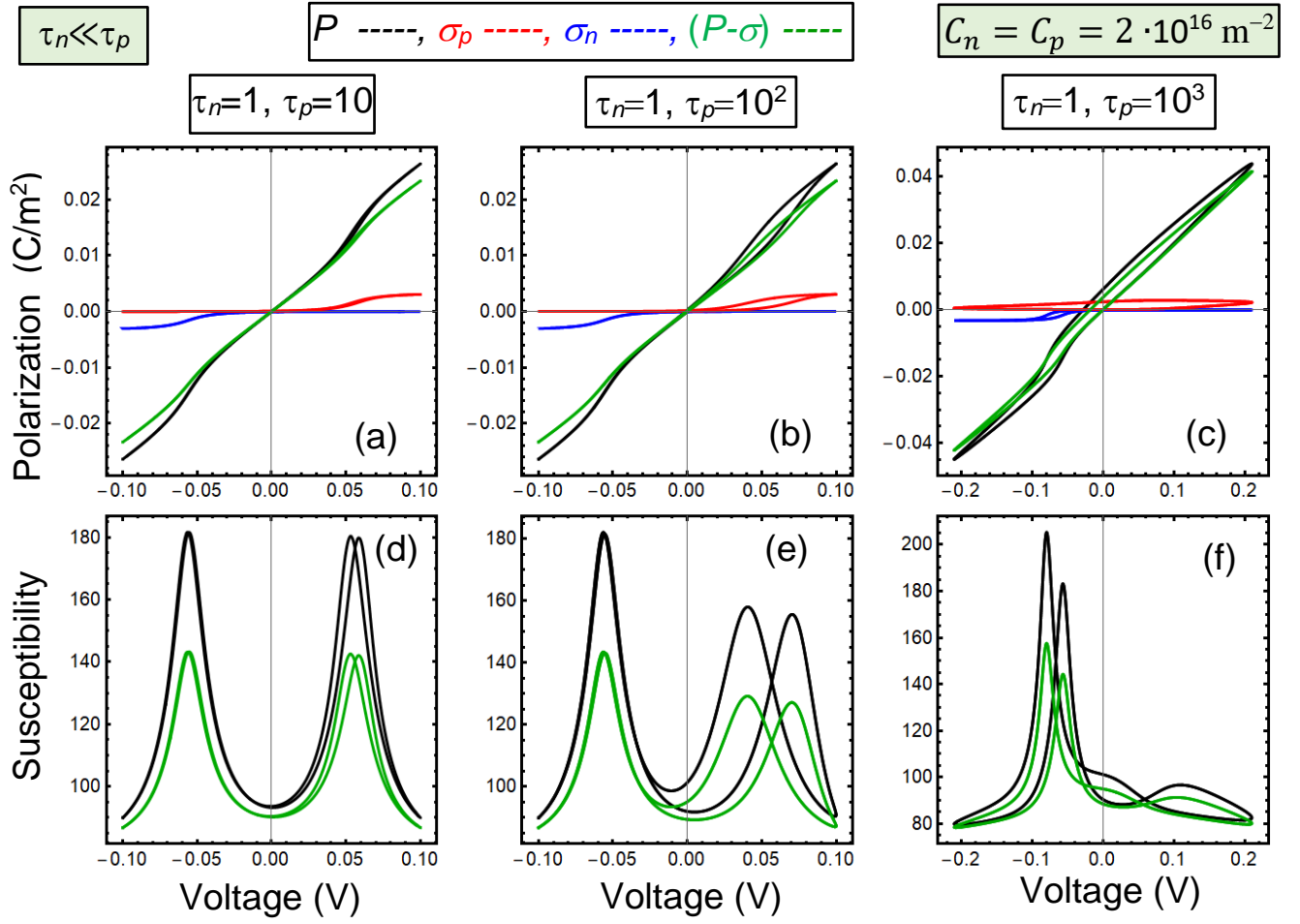

**FIGURE 8. Asymmetric paraelectric-like antiferroionic states. (a, b, c)** The quasi-static voltage dependences of the polarization $\bar{P}(V)$ (black curves), positive $\sigma_p(V)$ (red curves) and negative $\sigma_n(V)$ (blue curves) surface charges, and the difference $\bar{P}(V) - \sigma(V)$ (green curves). **(d, e, f)** The quasi-static dependences of the dielectric susceptibility $\frac{\partial \bar{P}}{\partial V}$ (black curves) and effective capacitance $\frac{\partial}{\partial V}(\bar{P} - \sigma)$ (green curves). The dependences are calculated for several values of relaxation times $\tau_n = 1, \tau_p = 10$ **(a, d)**, $\tau_n = 1, \tau_p = 10^2$ **(b, e)**, and $\tau_n = 1, \tau_p = 10^3$ **(c, f)** in the units of $\tau_{Kh}$. Other parameters: $h_f = 4$ nm, $w = 8$ nm, $d_1 = 0.4$ nm, $d_2 = 1.2$ nm, $\varepsilon_d = 100$, $\varepsilon_u = 10$, $u_m = +1\%$, and $T = 298$ K, $Z_p = -Z_n = +1$, $g_p = g_n = 1$, $\Delta G_p^0 = \Delta G_n^0 = 0.1$ eV, and $C_n = C_p = 2 \cdot 10^{16}$ m$^{-2}$.

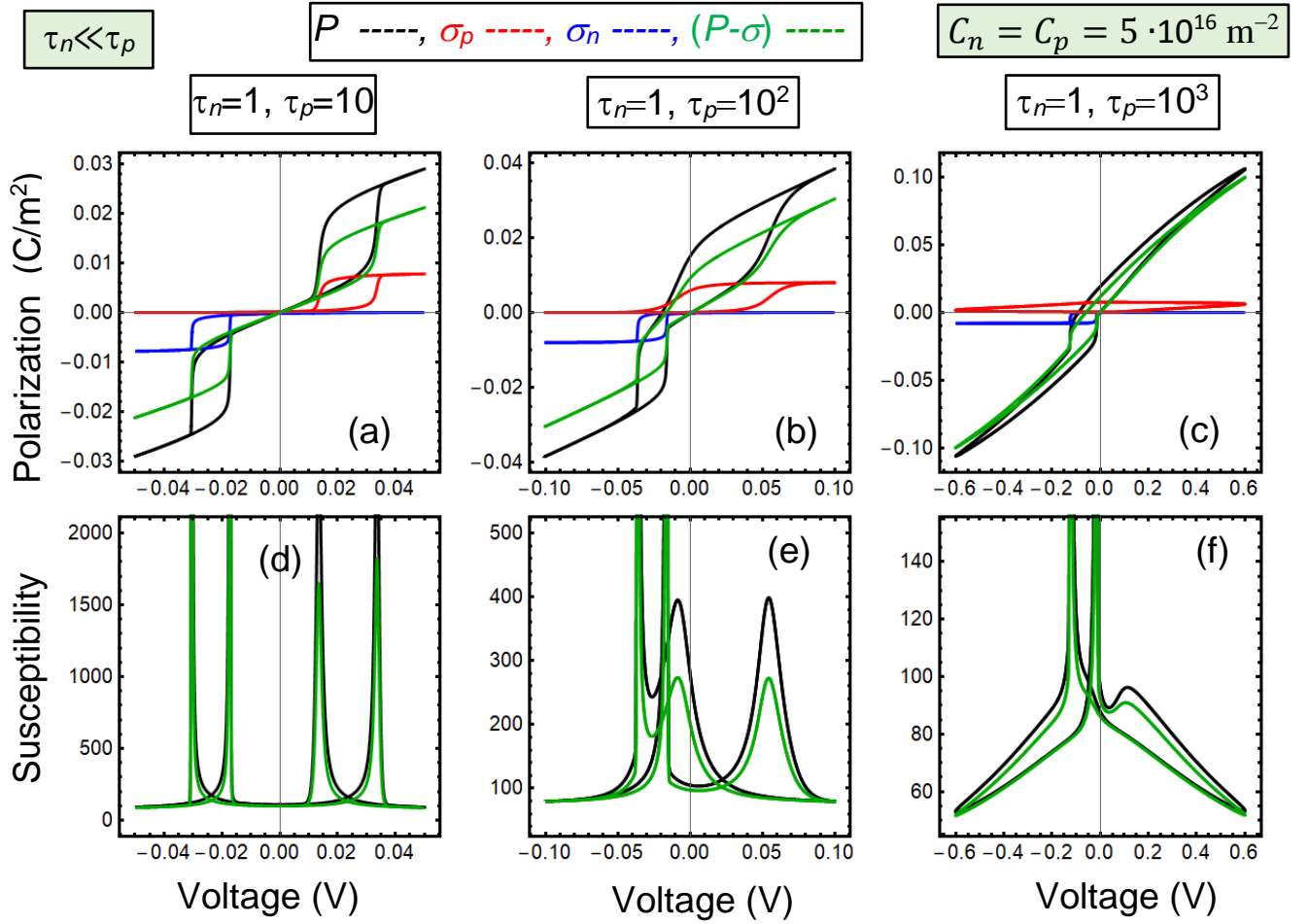

**FIGURE 9. Asymmetric antiferroionic states.** (**a, b, c**) The quasi-static voltage dependences of the polarization $\bar{P}(V)$ (black curves), positive $\sigma_p(V)$ (red curves) and negative $\sigma_n(V)$ (blue curves) surface charges, and the difference $\bar{P}(V) - \sigma(V)$ (green curves). (**d, e, f**) The quasi-static dependences of the dielectric susceptibility $\frac{\partial \bar{P}}{\partial V}$ (black curves) and effective capacitance $\frac{\partial}{\partial V}(\bar{P} - \sigma)$ (green curves). The dependences are calculated for the surface charge concentrations $C_n = C_p = 5 \cdot 10^{16}$ m$^{-2}$ and several values of relaxation times $\tau_n = 1, \tau_p = 10$ (**a, d**), $\tau_n = 1, \tau_p = 10^2$ (**b, e**), and $\tau_n = 1, \tau_p = 10^3$ (**c, f**) in the units of $\tau_{Kh}$. Other parameters are the same as in **Fig. 8**.

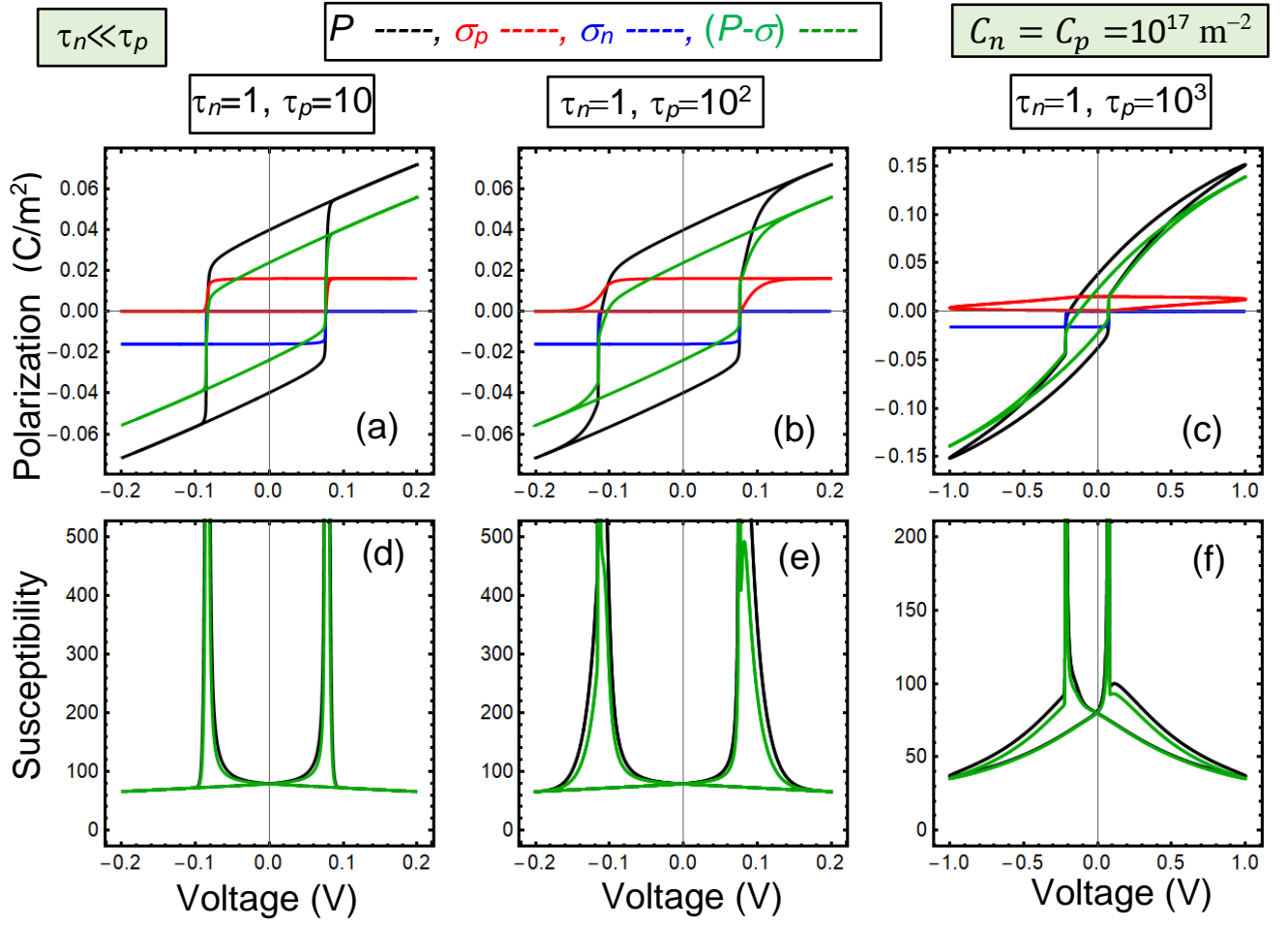

**FIGURE 10. Asymmetric ferroionic states. (a, b, c)** The quasi-static voltage dependences of the polarization $\bar{P}(V)$ (black curves), positive $\sigma_p(V)$ (red curves) and negative $\sigma_n(V)$ (blue curves) surface charges, and the difference $\bar{P}(V) - \sigma(V)$ (green curves). **(d, e, f)** The quasi-static dependences of the dielectric susceptibility $\frac{\partial \bar{P}}{\partial V}$ (black curves) and effective capacitance $\frac{\partial}{\partial V}(\bar{P} - \sigma)$ (green curves). The dependences are calculated for the surface charge concentrations $C_n = C_p = 10^{17}$ m$^{-2}$ and several values of relaxation times $\tau_n = 1, \tau_p = 10$ **(a, d)**, $\tau_n = 1, \tau_p = 10^2$ **(b, e)**, and $\tau_n = 1, \tau_p = 10^3$ **(c, f)** in the units of $\tau_{Kh}$. Other parameters are the same as in **Fig. 8**.

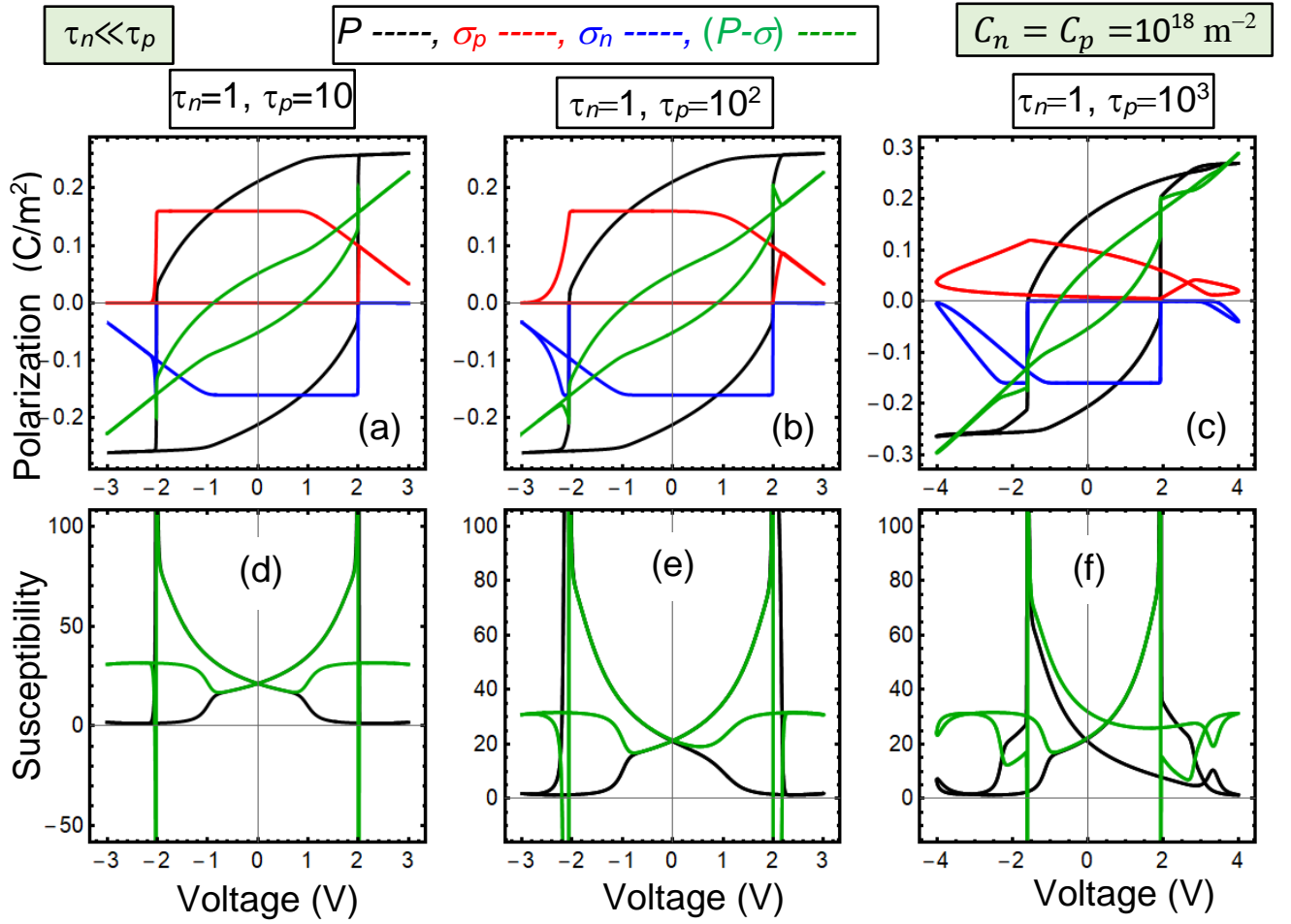

**FIGURE 11. Almost symmetric ferroelectric-like ferroionic states. (a, b, c)** The quasi-static voltage dependences of the polarization $\bar{P}(V)$ (black curves), positive $\sigma_p(V)$ (red curves) and negative $\sigma_n(V)$ (blue curves) surface charges, and the difference $\bar{P}(V) - \sigma(V)$ (green curves). **(d, e, f)** The quasi-static dependences of the dielectric susceptibility $\frac{\partial \bar{P}}{\partial V}$ (black curves) and effective capacitance $\frac{\partial}{\partial V}(\bar{P} - \sigma)$ (green curves). The dependences are calculated for the surface charge concentrations $C_n = C_p = 10^{18}$ m$^{-2}$ and several values of relaxation times $\tau_n = 1, \tau_p = 10$ **(a, d)**, $\tau_n = 1, \tau_p = 10^2$ **(b, e)**, and $\tau_n = 1, \tau_p = 10^3$ **(c, f)** in the units of $\tau_{Kh}$. Other parameters are the same as in **Fig. 8**.

### C. The role of the size effect on polarization and susceptibility hysteresis loops

**Figures 12** and **13** illustrate the influence of the size effect (i.e., core thickness $h_f$) on polarization and susceptibility hysteresis loops for the cases $\tau_n = \tau_p$ and $\tau_n \ll \tau_p$. The top rows show quasi-static voltage dependences of the polarization $\bar{P}(V)$, positive and negative surface charges, $\sigma_p(V)$ and $\sigma_n(V)$, and the difference $\bar{P}(V) - \sigma(V)$. The overall peculiarity of the figures is the transition from the AFE-type double hysteresis loop to the FE-type single loop with an increase of $h_f$. However, the quasi-static dependences of the dielectric susceptibility, $\frac{\partial \bar{P}}{\partial V}$, and effective capacitance, $\frac{\partial}{\partial V}(\bar{P} - \sigma)$, still have four symmetric (for $\tau_n = \tau_p$) or four asymmetric (for $\tau_n \ll \tau_p$) maxima indicating AFEI or mixed AFEI-FEI states, respectively. In particular, **Fig. 12** illustrates the size-induced transition from symmetric AFEI to FEI states; and **Fig. 13** illustrates the transition from asymmetric AFEI to FEI states with an $h_f$ increase from 3 nm to 5 nm. One of the most interesting behaviors is the strongly pinched polarization loops and four well-separated maxima of the dielectric susceptibility observed in a 3-nm SPS core, shown in **Figs.11(a)** and **11(d)**, **12(a)** and **12(d)**. The loop pinching in ultra-small ferroelectric nanoparticles, which should be paraelectric under linear screening conditions, is attributed to nonlinear electronic-ionic screening. The effect, in principle, can be an additional contribution for the explanation of the unusual multiple maxima of electric current observed in ferroelectric nanoparticles in a nonpolar fluid suspension, which were attributed to aggregation and disaggregation of multipole particles in an AC field [45].

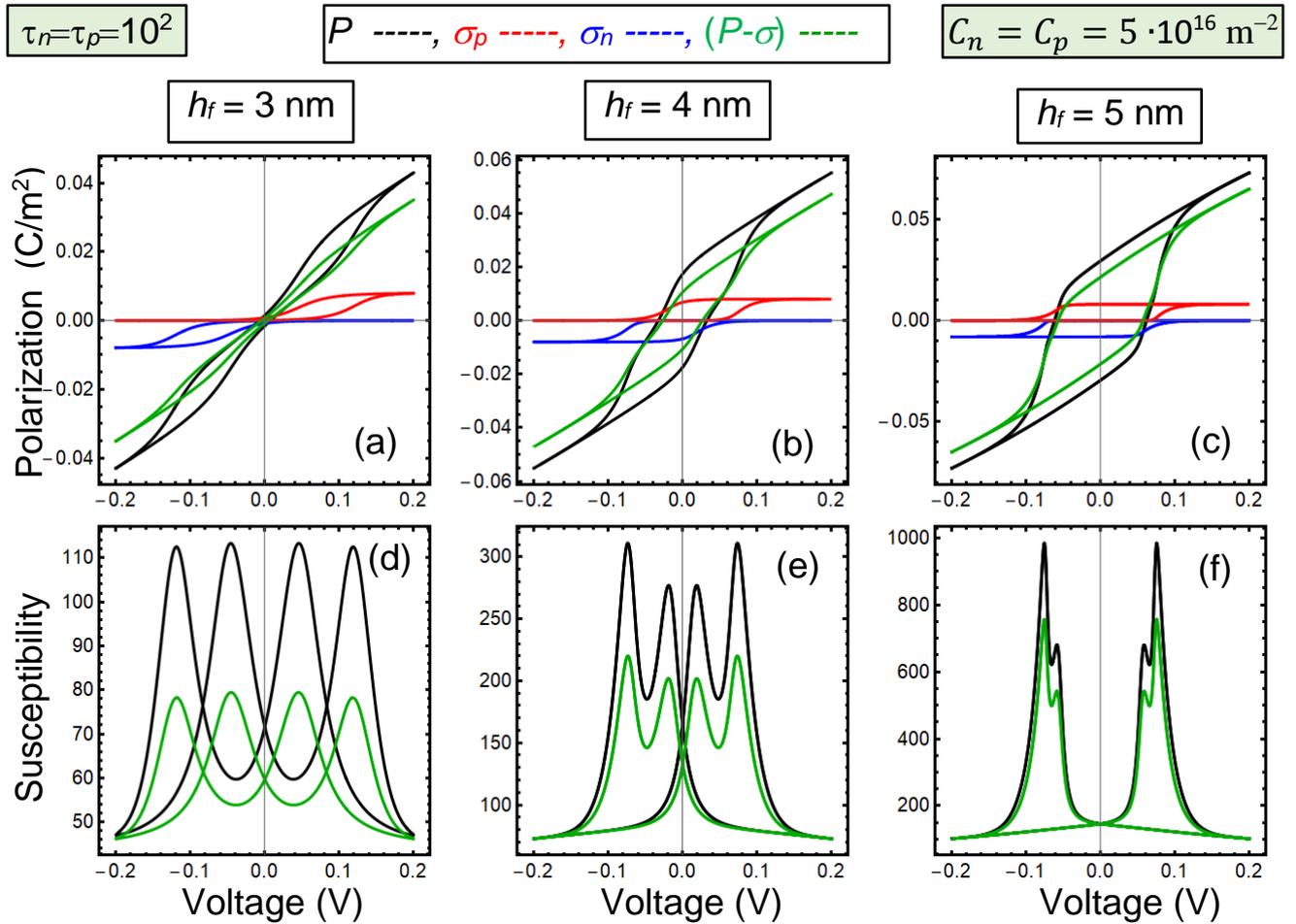

**FIGURE 12. Size-induced transition from symmetric antiferrionic to ferroionic states.** (a, b, c) The quasi-static voltage dependences of the polarization $\bar{P}(V)$ (black curves), positive $\sigma_p(V)$ (red curves) and negative $\sigma_n(V)$ (blue curves) surface charges, and the difference $\bar{P}(V) - \sigma(V)$ (green curves). (d, e, f) The quasi-static dependences of the dielectric susceptibility $\frac{\partial \bar{P}}{\partial V}$ (black curves) and effective capacitance $\frac{\partial}{\partial V}(\bar{P} - \sigma)$ (green curves). The dependences are calculated for several thickness of the SPS core $h_f = 3$ nm (**a, d**), $h_f = 4$ nm (**b, e**), and $h_f = 5$ nm (**c, f**). Other parameters: $w = 8$ nm, $d_1 = 0.4$ nm, $d_2 = 1.2$ nm, $\varepsilon_l = 100$, $\varepsilon_u = 10$, $u_m = +1\%$, and $T = 298$ K, $Z_p = -Z_n = +1$, $g_p = g_n = 1$, $\Delta G_p^0 = \Delta G_n^0 = 0.1$ eV, $C_n = C_p = 5 \cdot 10^{16}$ m$^{-2}$ and $\tau_n = \tau_p = 10^2 \tau_{Kh}$.

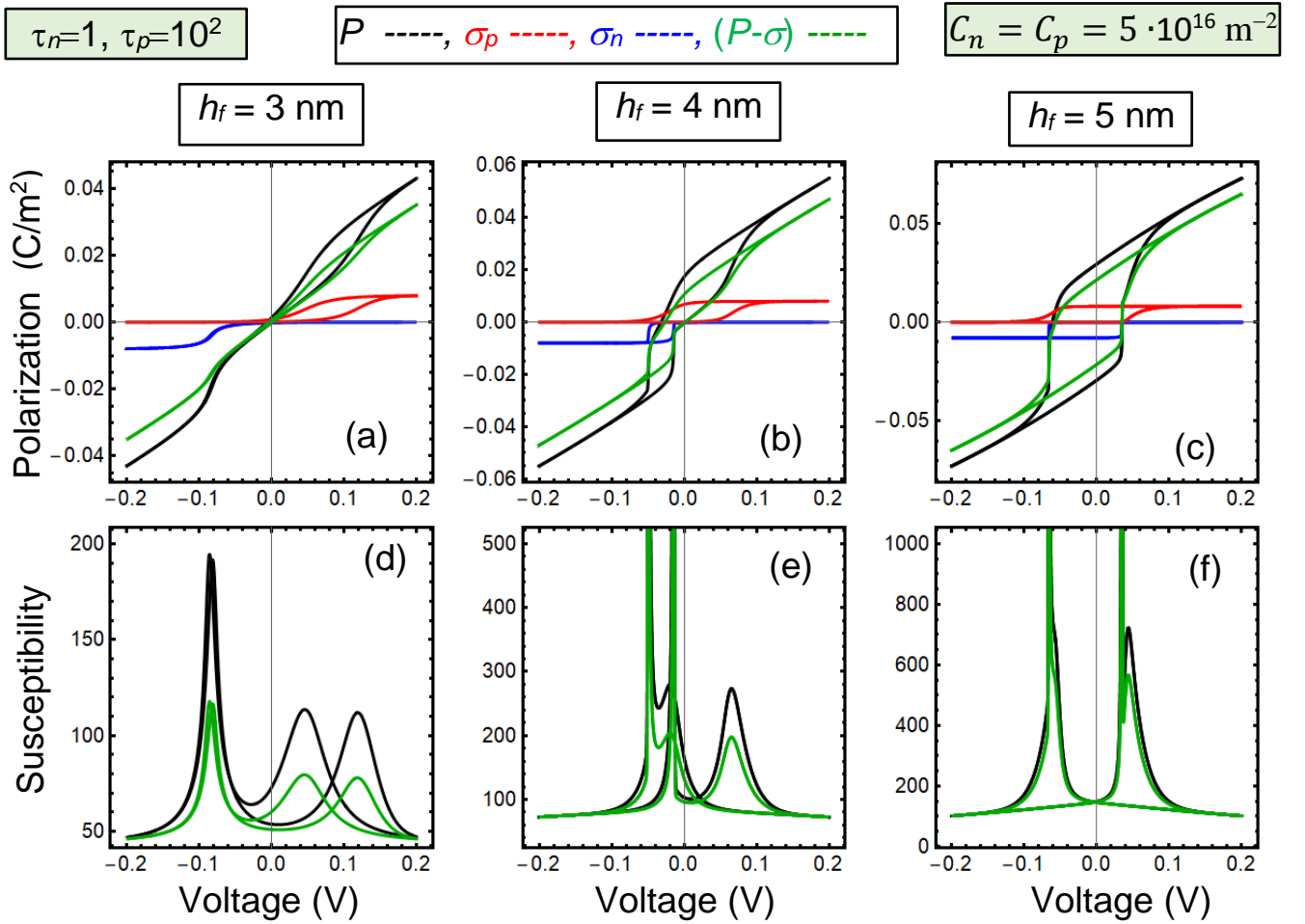

**FIGURE 13. Size-induced transition from asymmetric antiferrionic to ferroionic states.** (**a, b, c**) The quasi-static voltage dependences of the polarization $\bar{P}(V)$ (black curves), positive $\sigma_p(V)$ (red curves) and negative $\sigma_n(V)$ (blue curves) surface charges, and the difference $\bar{P}(V) - \sigma(V)$ (green curves). (**d, e, f**) The quasi-static dependences of the dielectric susceptibility $\frac{\partial \bar{P}}{\partial V}$ (black curves) and effective capacitance $\frac{\partial}{\partial V}(\bar{P} - \sigma)$ (green curves). The dependences are calculated for several thickness of the SPS core $h_f = 3$ nm (**a, d**), $h_f = 4$ nm (**b, e**), and $h_f = 5$ nm (**c, f**). Relaxation times $\tau_n = 1\tau_{Kh}$ and $\tau_p = 10^2 \tau_{Kh}$. Other parameters are the same as in **Fig. 12**.

## V. CONCLUSIONS

The polar states of uniaxial ferroelectric nanoparticles interacting with the surface system of electronic and ionic charges with a broad distribution of mobilities is explored, corresponding to the experimental case of nanoparticles in solution or ambient conditions. Due to the nonlinear electric interaction between the ferroelectric dipoles and surface charges with slow relaxation dynamics in an external field, the transitions between the PE-like, AFEI, and FE-like FEI states emerge.

The polarization and susceptibility in these systems is very sensitive to the concentrations, formation energies, and relaxation times of the screening charges. By increasing the surface charge concentrations one can continuously switch the state of ferroelectric core between PE-like, AFEI, mixed AFEI-FEI, FEI, and FE-like FEI states, for which we establish and analyzed the distinct features of polarization, surface charges, and dielectric susceptibility dynamic response to a periodic external field. In particular, the voltage dependences of the polarization and negative and positive surface charges are quasilinear and hysteresis-less in the PE-like state, which exists for small concentrations of surface charges. When the concentrations increase an antiferroelectric-type double hysteresis loop appears. The voltage positions of the double loops opening almost coincide with the opening positions of the positive and negative charges' "minor" loops, which are well-separated. For the minor loops, the surface charge opens at some critical voltages due to the strongly nonlinear exponential dependence of their density on the electric overpotential. This behavior corresponds to the AFEI state induced in the ferroelectric core by the interaction between ferroelectric dipoles and surface screening charges. Further increase of the screening charge leads to the appearance of the pinched polarization loops. The behavior corresponds to the mixed AFEI-FEI state in the ferroelectric core. The FEI state of the core, which is characterized by a single FE-type hysteresis loop, is induced by screening charges in higher concentrations. The loops in the FEI state become seemingly indistinguishable from the FE loops at high concentrations. However, this is an apparent effect only, because it is the FE-like FEI state supported by the nonlinear dynamics of surface charge, and the state does not exist without the nonlinear screening.

The simultaneous decrease of the positive and negative surface charges formation energies leads to the continuous transition from the AFEI to the FEI state in the ferroelectric core. The decrease of one of the formation energies, with the other fixed, leads to the mixing of AFEI and FEI states**.** The hysteresis loops in the mixed AFEI-FEI state are typically strongly asymmetric, horizontally shifted, distorted, and can be strongly pinched. Since the antiferroelectric-like double loops and pinched loops of polarization are often observed in polydomain ferroelectric thin films, as well as the fact that they are typically related with polydomain or vortex-like domain states in ferroelectric nanoparticles, their appearance in a single-domain ferroelectric core covered with ionic-electronic screening charges seems unusual and requires further studies. In the considered case, the appearance of the antiferroelectric-like loops is caused by the minor loops of the surface charges, which are absent at small voltages and open above the critical voltage.

The above-described average polarization and susceptibility dependences on the applied voltage are symmetric for equal relaxation times of the positive and negative surface charges, and becomes strongly asymmetric when these times differ by one or more orders of magnitude. The asymmetry originates from the strong retardation of the screening by one type of charge carrier with respect to the other~~, as well as the period of applied voltage~~. The retardation of dynamical screening is responsible for both the asymmetry and the significant horizontal shift of the polarization and surface charge loops, as well as the disappearance of double AFEI loops. Thus, we conclude, that the crossover between different polar states can be controlled not only by the static characteristics of the surface charges, such as concentrations and formation energies, but also by their relaxation dynamics in an applied field.

**Acknowledgements.** A.N.M. acknowledges EOARD project 9IOE063 and related STCU partner project P751a. E.A.E. acknowledges CNMS2021-B-00843 "Effect of surface ionic screening on polarization reversal scenario in antiferroelectric thin films: analytical theory, machine learning, PFM and cKPFM experiments". This effort (S.V.K) was supported as part of the center for 3D Ferroelectric Microelectronics (3DFeM), an Energy Frontier Research Center funded by the U.S. Department of Energy (DOE), Office of Science, Basic Energy Sciences under Award Number DE-SC0021118, and performed in part at the Oak Ridge National Laboratory's Center for Nanophase Materials Sciences (CNMS), a U.S. Department of Energy, Office of Science User Facility.

**Authors' contribution.** A.N.M. and S.V.K. generated the research idea, analyzed results and wrote the manuscript draft. A.N.M. formulated the problem, jointly with M.Ye. performed analytical calculations and prepared figures. E.A.E. wrote the codes. D.R.E. worked on the results explanation and manuscript improvement. All co-authors discussed the results.

# SUPPLEMENT

## Appendix A

Let us consider a capacitor of thickness $h$ filled with a polarized ferroelectric film of thickness $h_f$ sandwiched between two dielectric layers of thickness $d_1$ and $d_2$, as shown in **Fig. 1**. Note that $h_f = h - d_1 - d_2$.

The equations and boundary conditions are the Laplace equation for the electric potential $\varphi$ that is fixed at the conducting electrodes. The connection between the electric displacement $\vec{D}$ and the electric field, $\vec{E} = -\nabla\varphi$, in the ferroelectric ($f$) and the dielectric ($u$, $l$) layers are:

$$\vec{D}^f = \varepsilon_0\varepsilon_b\vec{E}^f + \vec{P}, \quad \vec{D}^u = \varepsilon_0\varepsilon_u\vec{E}^u, \quad \vec{D}^l = \varepsilon_0\varepsilon_d\vec{E}^l, \tag{A.1}$$

where $\varepsilon_0$ is a universal dielectric constant, $\varepsilon_b$ is a relative background permittivity of ferroelectric, $\varepsilon_u$ and $\varepsilon_l$ are relative dielectric permittivity of the dielectric layers, with the superscripts " $u$ " and " $l$ " denoting the upper and lower dielectric layers. Since $\text{div}\,\vec{D} = 0$ under the absence of free space charges, and in the case when all variables depend on the $z$-coordinate and the ferroelectric polarization $\vec{P}$ is directed along the polar axis $z$, $\vec{P} = (0,0,P)$, we obtain the following electrostatic equations in the ferroelectric and dielectric layers:

$$\varepsilon_0\varepsilon_b \frac{\partial^2 \varphi_f}{\partial z^2} = \frac{\partial P}{\partial z}, \quad \varepsilon_0\varepsilon_u \frac{\partial^2 \varphi_u}{\partial z^2} = 0, \quad \varepsilon_0\varepsilon_l \frac{\partial^2 \varphi_l(z)}{\partial z^2} = 0. \tag{A.2}$$

The tangential component of the electric field is homogenous at the interfaces, and the normal components of the displacement differ by the value of the surface charge densities $\sigma_1$ and $\sigma_2$ at the interfaces. Taking this into account, the boundary conditions are as follows:

$$\varphi_l(d_1) - \varphi_f(d_1) = 0, \tag{A.3a}$$

$$D_z^l(d_1) - D_z^f(d_1) = -\sigma_1, \tag{A.3b}$$

$$\varphi_u(h - d_2) - \varphi_f(h - d_2) = 0, \tag{A.3c}$$

$$D_z^u(h - d_2) - D_z^f(h - d_2) = -\sigma_2, \tag{A.3d}$$

$$\varphi_l(0) = 0, \quad \varphi_u(h) = 0. \tag{A.3e}$$

Using the superposition principle, at first, we can consider the absence of the surface charges ($\sigma_1 = \sigma_2 = 0$). In the case of zero surface charge, the general solution of Eqs.(A.2) is:

$$\varphi_f(z) = \frac{A_f}{\varepsilon_0\varepsilon_b} - \frac{1}{\varepsilon_0\varepsilon_b}\int_z^{h-d_2} P(\tilde{z})d\tilde{z} + \frac{z}{\varepsilon_0\varepsilon_b}C_f, \tag{A.4a}$$

$$\varphi_l(z) = \frac{C_l z}{\varepsilon_0\varepsilon_l} + \frac{A_l}{\varepsilon_0\varepsilon_l}, \quad \varphi_u(z) = \frac{C_u z}{\varepsilon_0\varepsilon_u} + \frac{A_u}{\varepsilon_0\varepsilon_u}. \tag{A.4b}$$

After the substitution of the expressions (A.4) in the boundary conditions (A.3) and using expressions (A.1), the elementary transformations lead to the solution:

$$\varphi_f(z) = -\frac{1}{\varepsilon_0 \varepsilon_b} \int_z^{h-d_2} P(\tilde{z}) d\tilde{z} - \frac{z}{\varepsilon_0 \varepsilon_b} \frac{\langle P \rangle (h_f/\varepsilon_b)}{d_1\left(\frac{1}{\varepsilon_l} - \frac{1}{\varepsilon_b}\right) + \left(\frac{d_2}{\varepsilon_u} + \frac{h-d_2}{\varepsilon_b}\right)} + \frac{1}{\varepsilon_0 \varepsilon_b}\left(\frac{1}{\varepsilon_u} - \frac{1}{\varepsilon_b}\right) \frac{\langle P \rangle h_f d_2}{d_1\left(\frac{1}{\varepsilon_l} - \frac{1}{\varepsilon_b}\right) + \left(\frac{d_2}{\varepsilon_u} + \frac{h-d_2}{\varepsilon_b}\right)},$$
(A.5a)

$$\varphi_l(z) = -\frac{z}{\varepsilon_0 \varepsilon_d} \frac{1}{\varepsilon_b} \frac{h_f \langle P \rangle}{d_1\left(\frac{1}{\varepsilon_l} - \frac{1}{\varepsilon_b}\right) + \left(\frac{d_2}{\varepsilon_u} + \frac{h-d_2}{\varepsilon_b}\right)},$$
(A.5b)

$$\varphi_u(z) = -\frac{1}{\varepsilon_0 \varepsilon_u} \frac{z}{\varepsilon_b} \frac{h_f \langle P \rangle}{d_1\left(\frac{1}{\varepsilon_l} - \frac{1}{\varepsilon_b}\right) + \left(\frac{d_2}{\varepsilon_u} + \frac{h-d_2}{\varepsilon_b}\right)} + \frac{1}{\varepsilon_0 \varepsilon_u} \frac{h}{\varepsilon_b} \frac{h_f \langle P \rangle}{d_1\left(\frac{1}{\varepsilon_l} - \frac{1}{\varepsilon_b}\right) + \left(\frac{d_2}{\varepsilon_u} + \frac{h-d_2}{\varepsilon_b}\right)}.$$
(A.5c)

Hereinafter $\langle P \rangle = \frac{1}{h_f} \int_{d_1}^{h-d_2} P(\tilde{z}) d\tilde{z}$.

The fields are simply the derivatives with a negative sign:

$$E_z^f(z) = \frac{1}{\varepsilon_0 \varepsilon_b}\left[\langle P \rangle \frac{\frac{h_f}{\varepsilon_b}}{\left(\frac{d_1}{\varepsilon_l} + \frac{d_2}{\varepsilon_u}\right) + \frac{h_f}{\varepsilon_b}} - P(z)\right],$$
(A.6a)

$$E_z^l(z) = \frac{\langle P \rangle}{\varepsilon_0 \varepsilon_d} \frac{\frac{h_f}{\varepsilon_b}}{\left(\frac{d_1}{\varepsilon_l} + \frac{d_2}{\varepsilon_u}\right) + \frac{h_f}{\varepsilon_b}}, \qquad E_z^u(z) = \frac{\langle P \rangle}{\varepsilon_0 \varepsilon_u} \frac{\frac{h_f}{\varepsilon_b}}{\left(\frac{d_1}{\varepsilon_l} + \frac{d_2}{\varepsilon_u}\right) + \frac{h_f}{\varepsilon_b}}.$$
(A.6b)

For a particular case of constant polarization $E_z^f = \frac{\langle P \rangle}{\varepsilon_0 \varepsilon_b} \frac{\frac{d_1}{\varepsilon_l} + \frac{d_2}{\varepsilon_u}}{\left(\frac{d_1}{\varepsilon_l} + \frac{d_2}{\varepsilon_u}\right) + \frac{h_f}{\varepsilon_b}}$. Without dead layers ($d_{1,2} \to 0$), we get the following solution:

$$E_z^f(z) = -\frac{1}{\varepsilon_0 \varepsilon_b}[P(z) - \langle P \rangle], \qquad E_z^d(z) = \frac{\langle P \rangle}{\varepsilon_0 \varepsilon_l}, \qquad E_z^u(z) = \frac{\langle P \rangle}{\varepsilon_0 \varepsilon_u}.$$
(A.7a)

**Table SI.** The parameters for a bulk ferroelectric $Sn_2P_2S_6$

| Parameter | Dimension | Values for $Sn_2P_2S_6$ collected from Refs. [64, 65, 66] |
|---|---|---|
| $\varepsilon_b$ | 1 | 7 [*] |
| $\alpha_T$ | m/F | $1.44 \times 10^6$ |
| $T_C$ | K | 337 |
| $\beta$ | $C^{-4} \cdot m^5 J$ | $9.40 \times 10^8$ |
| $\gamma$ | $C^{-6} \cdot m^9 J$ | $5.11 \times 10^{10}$ |
| $g_{ij}$ | $m^3/F$ | $g_{11} = 5.0 \times 10^{-10}$ [**], $g_{44} = 2.0 \times 10^{-10}$ |
| $s_{ij}$ | 1/Pa | $s_{11} = 4.1 \times 10^{-12}$, $s_{12} = -1.2 \times 10^{-12}$, $s_{44} = 5.0 \times 10^{-12}$ |
| $Q_{ij}$ | $m^4/C^2$ | $Q_{11} = 0.22$, $Q_{12} = 0.12$, $Q_{12} \approx Q_{13} \approx Q_{23}$ [****] |
| $F_{ij}$ | $m^3/C$ | $F_{11} = 1.0 \times 10^{-11}$, $F_{12} = 0.9 \times 10^{-11}$, $F_{44} = 3 \times 10^{-11}$ |

[*] estimated from a refraction index value

[**] the order of magnitude is estimated from the uncharged domain wall width [64-66]

[***] estimation is based on the values of the spontaneous polarization and permittivity at room temperature

[****] estimation of electrostriction is based on thermal expansion data from Say et al.[66].